\documentclass[preprint,12pt,authoryear]{elsarticle}
\usepackage{amssymb}
\usepackage{amsmath}
\usepackage{cite}
\usepackage{amsmath,amssymb,amsfonts}
\usepackage{adjustbox}
\usepackage{mdframed}
\usepackage{algorithm}
\usepackage{algorithmicx}
\usepackage{algpseudocode}
\usepackage{caption}
\usepackage{subcaption}
\usepackage{multicol}
\usepackage{import}
\usepackage{balance}
\usepackage{float}
\usepackage{graphicx}
\usepackage{enumitem}
\usepackage{array}
\usepackage{tabularx}
\usepackage{textcomp}
\usepackage{xcolor}
\usepackage{xspace}
\usepackage{booktabs}
\usepackage{afterpage}
\usepackage{listings}
\usepackage{makecell}
\usepackage{upquote}
\usepackage{lineno}
\usepackage[hyphens]{url}
\usepackage{hyperref}
\usepackage{cleveref}

\newtheorem{definition}{Definition}

\definecolor{light-gray}{gray}{0.95}
\definecolor{light-gray-bg}{gray}{0.9}
\definecolor{pgreen}{RGB}{5,205,107}
\definecolor{pblue}{RGB}{2,154,223}
\newcolumntype{Y}{>{\centering\arraybackslash}X}
\newcolumntype{L}{>{\centering\arraybackslash}b{2.5cm}}
\newcolumntype{M}{>{\centering\arraybackslash}b{2.5cm}}
\newcolumntype{N}{>{\centering\arraybackslash}b{1.5cm}}
\newcolumntype{O}{>{\centering\arraybackslash}b{1cm}}
\newcolumntype{P}{>{\centering\arraybackslash}b{2.5cm}}

\lstdefinelanguage{JavaScript}{
  keywords={break, case, catch, continue, debugger, delete, do, else, finally, for, function, if, in, typeof, instanceof, new, return, switch, this, throw, try, void, var, while, with, require, module, exports},
  morecomment=[l]{//},
  morecomment=[s]{/*}{*/},
  morestring=[b]',
  morestring=[b]",
  ndkeywords={class, export, boolean, throw, implements, import, this},
  keywordstyle=\color{blue}\bfseries,
  ndkeywordstyle=\color{darkgray}\bfseries,
  identifierstyle=\color{black},
  commentstyle=\color{purple}\ttfamily,
  sensitive=true
}

\newmdenv[
  backgroundcolor=light-gray,
  linecolor=black,
  linewidth=1pt,
  innerleftmargin=8pt,
  innerrightmargin=8pt,
  innertopmargin=8pt,
  innerbottommargin=8pt,
  roundcorner=10pt,
  nobreak=true
]{answerbox}

\newcommand{\lstbg}[3][0pt]{{\fboxsep#1\colorbox{#2}{\strut #3}}}
\definecolor{codegreen}{rgb}{0,0.6,0}
\lstdefinelanguage{diff}{
    basicstyle=\ttfamily\footnotesize,
	morecomment=[f][\color{red}]{---}, 
	morecomment=[f][\color{codegreen}]{+++},
	morecomment=[f][\lstbg{red!20}]{-\ },
	morecomment=[f][\lstbg{green!20}]{\ +\ },
	morecomment=[f][\color{blue}]{@@},}
\makeatletter
\lst@AddToHook{OnEmptyLine}{\addtocounter{lstnumber}{-1}}
\makeatother
\newcommand{\edit}[1]{\textcolor{black}{#1}}

\newcommand{\javascript}{JavaScript\xspace}
\newcommand{\nodejs}{Node.js\xspace}
\newcommand{\npm}{npm\xspace}
\newcommand{\package}{\emph{package.json}\xspace}
\newcommand{\packagelock}{\emph{package-lock.json}\xspace}
\newcommand{\nodemodules}{\texttt{node\_modules}\xspace}
\newcommand{\stubbifier}{\texttt{Stubbifier}\xspace}
\newcommand{\depCheck}{\texttt{depcheck}\xspace}
\newcommand{\depPrune}{\texttt{DepPrune}\xspace}
\newcommand{\commonjs}{CommonJS\xspace}
\newcommand{\require}{\texttt{require()}}
\newcommand{\nbPackages}{91\xspace}
\newcommand{\nbPackagesBloated}{90\xspace}
\newcommand{\nbPackagesStubbifier}{77\xspace}
\newcommand{\nbPackagesDirectBloated}{46\xspace}
\newcommand{\totalTestCases}{10,268\xspace}
\newcommand{\totalNbDeps}{50,488\xspace}
\newcommand{\totalNbDirectDeps}{869\xspace}
\newcommand{\totalNbIndirectDeps}{49,619\xspace}

\newcommand{\totalNbNonBloatedDeps}{24,972\xspace}
\newcommand{\ratioBloatedDeps}{50.6\%\xspace}
\newcommand{\ratioDirectBloatedDeps}{13.8\%\xspace}
\newcommand{\ratioIndirectBloatedDeps}{51.3\%\xspace}
\newcommand{\totalNbBloatedDeps}{25,566\xspace}
\newcommand{\totalNbBloatedDirectDeps}{120\xspace}
\newcommand{\totalNbBloatedIndirectDeps}{25,446\xspace}
\newcommand{\totalNbBloatedInrectDepsByDirect}{4,167\xspace}
\newcommand{\testCov}{96.9\%\xspace}
\newcommand{\totalAccessedModules}{135,095\xspace}
\newcommand{\directBloatedByDepcheck}{162\xspace}

\newcommand{\subsettotalNbDeps}{21,614\xspace}
\newcommand{\subsettotalNbDirectDeps}{280\xspace}
\newcommand{\subsettotalNbIndirectDeps}{21,334\xspace}
\newcommand{\subsettotalNbBloatedDeps}{12,013\xspace}
\newcommand{\subsettotalNbBloatedDirectDeps}{30\xspace}

\newcommand{\subsetratioBloatedDeps}{55.6\%\xspace}
\newcommand{\subsetratioDirectBloatedDeps}{10.7\%\xspace}
\newcommand{\subsetratioIndirectBloatedDeps}{56.2\%\xspace}

\newcommand{\RQone}{What is the prevalence of bloated dependencies within the dependency trees of packages under our study?\xspace}
\newcommand{\RQtwo}{What impact does the removal of direct bloated dependencies have on the size of the dependency tree?\xspace}
\newcommand{\RQthree}{How does trace-based analysis compare to the state-of-the-art approaches for detecting \commonjs bloated dependencies? \xspace}

\journal{Journal of Systems and Software}

\begin{document}

\begin{frontmatter}

\title{Detecting and removing bloated dependencies in \commonjs packages} %% Article title

% \author{} %% Author name
\author{Yuxin Liu}
% \author{\fnm{Yuxin} \sur{Liu}}
% \email{yuxinli@kth.se}
\author{Deepika Tiwari}
% \author{\fnm{Deepika} \sur{Tiwari}}
% \email{deepikat@kth.se}
\author{Cristian Bogdan}
% \author{\fnm{Cristian} \sur{Bogdan}}
% \email{cristi@kth.se}
\author{Benoit Baudry}
% \author{\fnm{Benoit} \sur{Baudry}}
% \email{baudry@kth.se}

\affiliation{organization={KTH Royal Institute of Technology},%Department and Organization
            city={Stockholm},
            country={Sweden}}

%% Abstract
\begin{abstract}
\javascript packages are notoriously prone to bloat, a factor that significantly impacts the performance and maintainability of web applications.
While web bundlers and tree-shaking can mitigate this issue in client-side applications, state-of-the-art techniques have limitations on the detection and removal of bloat in server-side applications.
In this paper, we present the first study to investigate bloated dependencies within server-side \javascript applications, focusing on those built with the widely used and highly dynamic \commonjs module system.
We propose a trace-based dynamic analysis that monitors the OS file system, to determine which dependencies are not accessed during runtime.
To evaluate our approach, we curate an original dataset of \nbPackages \commonjs packages with a total of \totalNbDeps dependencies. 
Compared to the state-of-the-art dynamic and static approaches, our trace-based analysis demonstrates higher accuracy in detecting bloated dependencies.
Our analysis identifies \ratioBloatedDeps of the \totalNbDeps dependencies as bloated: \ratioDirectBloatedDeps of direct dependencies and \ratioIndirectBloatedDeps of indirect dependencies.
Furthermore, removing only the direct bloated dependencies by cleaning the dependency configuration file can remove a significant share of unnecessary bloated indirect dependencies while preserving function correctness.

\end{abstract}

%% Keywords

\end{frontmatter}

\section{Introduction}\label{sec:introduction}
Large-scale code reuse presents a valuable opportunity to speed up project development, building new features without reinventing the wheel \citep{krueger1992software, cox2019surviving, wittern2016look}.
However, reuse also artificially inflates the size of the codebase by pulling dependency code that is not actually needed by the project.
This unnecessary code, known as code bloat \citep{gkortzis2021software, soto2021comprehensive,PontaFPS21}, has a negative impact on maintenance \citep{weeraddana2024dependency, song2024efficiently, jafari2021dependency}, performance \citep{heo2018effective}, and security \citep{ChinthanetPPSKI20, PashchenkoPPSM22}.

Such bloat issues also affect code written in \javascript.
Its dominant usage is facilitated by the Node Package Manager (\npm\footnote{\href{https://www.npmjs.com}{https://www.npmjs.com}}), which provides reusable external packages (i.e., dependencies), but also contributes to code bloat. 
The portability of \javascript and the ease of integrating new \npm dependencies encourages fine-grained reuse, often resulting in the inclusion of small-sized dependencies, commonly referred to as ``trivial packages" (e.g., is-odd\footnote{\href{https://github.com/i-voted-for-trump/is-odd}{https://github.com/i-voted-for-trump/is-odd}}) \citep{abdalkareem2017developers,chowdhury2021untriviality}.
While this fine-grained modularity promotes reusability, it also amplifies the prevalence of dependency bloat.
Even small \javascript projects can have large dependency trees, \edit{which suffer from dependency overhead,} leading to maintenance and security challenges \citep{abdalkareem2020impact}.
\edit{For example, the 11-lines \texttt{left-pad} package, is reported to pose a significant threat to the software supply chain \citep{TrivialThreat}.}
Current research focuses primarily on mitigating code bloat at the function and file levels \citep{koishybayev2020mininode, turcotte2021stubbifier, vazquez2019slimming, treeshaking}. 
However, such fine-grained analyses do not support the total removal of unnecessary dependencies from the dependency specification file.
Consequently, these dependencies stay in the dependency tree and still need to be maintained and evolved even though they are not used in the project.

In this paper, we collect evidence on the prevalence of bloated dependencies within \npm packages, focusing on those using the widely adopted \commonjs module system.
Our focus on code bloat at the granularity of entire dependencies is twofold.
First, we aim to assist developers in managing dependencies when writing and maintaining their dependency specification files\footnote{\href{https://docs.npmjs.com/cli/v10/configuring-npm/package-json}{package.json, similar to pom.xml in Maven for Java}}.
%configuration files, such as \package.
Second, reducing bloat at the dependency level directly impacts the number of alarms raised by dependency bots \citep{rombaut2023there} and can reduce notification fatigue \citep{mirhosseini2017can}. To the best of our knowledge, this is the first study of code bloat at the granularity of npm packages.

Our study relies on dynamic analysis, due to the essential specifics of \commonjs and \javascript.
In \commonjs, developers access dependencies by invoking the \require\ function. 
The imperative nature of \require\ combined with the scripting nature of JavaScript pose significant challenges for tools aimed at researching or maintaining \commonjs packages. These challenges arise for several reasons: (i) as a function, \require\ may or may not be invoked depending on programming logic; (ii) as a function, \require\ can accept variable parameters, which (iii) may change from one code execution to the next; and (iv) \javascript's scripting nature allows \require\ calls to be generated dynamically as variable strings. As a result, static analysis methods that rely on examining \require\ calls within \javascript source code often produce imprecise results, as we shall demonstrate in this paper using the case of the standard \npm \depCheck\ tool.

To address the previously discussed challenges, we propose a trace-based analysis to identify bloated runtime dependencies, which we implement in a tool called \depPrune.
Our approach tracks interactions between the program and the file system at runtime to determine which dependencies are actually used.
We evaluate \depPrune against a novel dataset of open-source \commonjs packages, which we curate.
The \nbPackages packages in our dataset have a total of \totalNbDeps runtime dependencies.

Running \depPrune on our dataset reveals that a total of \totalNbBloatedDeps (\ratioBloatedDeps) dependencies are not accessed at runtime.
Notably, when \depPrune removes these dependencies from the packages in our dataset, the behavior of all packages is preserved, with a 100\% of the test cases still passing after the removal of the dependencies.
Developers can immediately act on direct bloated dependencies, by removing them from the dependency configuration file.
Two key insights from our study are that (i) more than half (50.5\%) of the \nbPackages packages have at least one direct bloated dependency, and (ii) removing a total of \totalNbBloatedDirectDeps direct bloated dependencies, also immediately removes \totalNbBloatedInrectDepsByDirect indirect dependencies.
As part of our study, we compare our trace-based technique of \depPrune, with the state-of-the-art dynamic technique based on test coverage \citep{turcotte2021stubbifier} and a static technique using a standard \npm tool \citep{depcheckgitaddress}.
Our comparative analysis reveals that \depPrune outperforms state-of-the-art techniques by achieving higher accuracy in identifying bloated dependencies, particularly in addressing dynamic execution patterns that challenge coverage-based and static methods.

Our primary contributions are outlined as follows:
\begin{itemize}[label= -]
    \item a novel dynamic analysis technique for detecting and removing bloated dependencies, implemented in a tool called \depPrune.
    \item empirical evidence highlighting the prevalence and impact of bloated dependencies within \commonjs packages.
    \item comparative analyses of state-of-the-art bloat detection techniques for server-side \javascript.
    \item a dataset of \nbPackages open-source \commonjs packages with strong test suites, accessible on Zenodo\footnote{\href{https://zenodo.org/doi/10.5281/zenodo.15090140}{10.5281/zenodo.15090140}}.
\end{itemize}

The subsequent sections of this paper are organized as follows: 
Section 2 introduces key concepts and terminology relevant to the dependency analysis of \commonjs packages.
Section 3 describes the procedure of our approach for analyzing bloated dependencies.
Section 4 outlines our research questions and explains the methodology to answer these questions.
Section 5 presents the results for each research question.
Section 6 delves into the threats to the validity of our study.
In Section 7, we discuss closely related work on debloating and dependency management.
Finally, Section 8 concludes the paper and discusses directions for future research.

%%=============================================================%%

%%=============================================================%%

%%=============================================================%%
\section{Background}\label{sec:background}
This section introduces the key concepts and terminology for dependency analysis of \nodejs packages that use the \commonjs module system and introduces the state-of-the-art static and dynamic debloating techniques.

\subsection{Structure of \npm Packaging System}
\label{subsec:npm-structure}

The dominant and widely used \npm repository manages reusable code in the form of \emph{packages}, which specify their dependencies in a declaration file called \package\footnote{\href{https://docs.npmjs.com/cli/v10/using-npm/package}{https://docs.npmjs.com/developers}}.
The dependencies declared in \package can be classified as runtime dependencies (required for execution) or development dependencies (used for testing, building, etc.).
A version range can be specified for each dependency. 
We refer to the dependencies declared in \package as \emph{direct dependencies}.

When working on a package, developers use the \npm tool to install all its dependencies via the \texttt{npm install} command. 
This process involves copying the \emph{source code} of the dependencies, which consists of \javascript \emph{modules} along with data files, e.g., in JSON format.
There are no binary formats involved, since \javascript is a scripting language (see \autoref{lst:scriptingrequire}).
Under the root of the package, the installed dependencies are stored in a directory named \texttt{node\_modules}, where each dependency is placed in its own folder named after the dependency.
In cases where multiple versions of the same dependency are required by different parts of the dependency tree, conflicting versions are resolved by installing each version in the \nodemodules folder of the respective parent dependency.
This ensures that each dependency gets the specific version it requires, even if this results in multiple versions of the same dependency being installed at different levels of the directory hierarchy.
This naming convention allows tools and developers to easily identify the dependency from its file path.

The installation automatically selects appropriate versions for all direct dependencies. 
Additionally, this process also retrieves all dependencies declared by the direct dependencies, which we refer to as \emph{indirect dependencies}.
The structure formed by direct and indirect dependencies is known as the \emph{dependency tree} of a package.
\npm records all the dependencies included in this tree in a file called \packagelock, which ensures consistency in future installations.
The \packagelock file documents the exact versions of dependencies that have been installed, and categorizes them based on their scope, such as \emph{runtime} (in \texttt{dependencies}) or \emph{development} (in \texttt{devDependencies}).

\subsection{\commonjs Module System and the Need for Dynamic Analysis}\label{subsec:commonjs-module-system}
\javascript originally lacked a built-in mechanism for referencing dependencies and code reuse.
Only with the emergence of \nodejs was a module system (the way to refer to dependencies) integrated into the language, \commonjs being the original way to package \javascript code for \nodejs \footnote{\url{https://nodejs.org/api/modules.html\#modules-commonjs-modules}}.
\commonjs has become the dominant module system in \nodejs applications, used by 
nearly 70\% of popular applications \citep{cjsdominant, cjsdominantgit}.

In \commonjs, modules reuse other modules by invoking the \require\ function and expose their functionalities to other modules by assigning the \texttt{exports} property of a special object called \texttt{module}.
\commonjs thus avoids introducing new language constructs, instead relying on existing ones like function calls and assignments.
For example, importing a dependency is done by a function call: \texttt{const importedObject = require(moduleName);}, and exporting one is done via an assignment: \texttt{module.exports = someObject}. 
However, since \require\ is a function, the exact modules imported are only known at runtime.
This introduces important challenges when trying to analyze \javascript code, e.g., debloating, necessitating the use of dynamic analysis \citep{turcotte2021stubbifier}.

Despite these challenges, we argue that it remains important to invest in improving the \commonjs system for several reasons.
First, \commonjs is the dominant module system for \nodejs, underpinning a significant portion of the \javascript projects.
Its widespread adoption highlights the practical value of developing methods to enhance its maintainability.
Second, the inherent flexibility of \commonjs, which allows developers to import dependencies conditionally and dynamically, is a double-edged sword: This flexibility enables diverse use cases and supports modularity, but it also complicates the detection of unused dependencies.
As a result, analyzing and addressing dependency bloat in \commonjs packages is particularly challenging yet crucial for improving dependency management.

Let us now revisit and elaborate on the four key challenges we identified in the introduction that make \commonjs particularly difficult for static analysis:

\emph{(i)\require\ is a function whose invocation depends on the programming logic. } Developers have the flexibility to use \require\ anywhere within their codebase.
%as \commonjs offers high versatility.
\autoref{lst:flexiblerequire} presents an excerpt from the package \texttt{inherits}, which is an indirect dependency of \texttt{airtap}.
In this snippet, the imported and exported modules vary according to runtime conditions.
This dynamic behavior allows modules to be loaded conditionally, making the dependency structure more adaptable to varying execution environments and use cases.

\emph{(ii) As a function, \require\ can accept variable parameters, (iii) which may change from one code execution to the next.}
It is common for developers to pass variables as parameters to the \require\ function.
In \autoref{lst:require_in_runtime}, this is illustrated by a plugin scenario, where the \texttt{pluginName} is computed in other parts of the code.
The \require\ call itself is part of a function that can be invoked repeatedly, potentially loading different dependency packages with each execution.

\lstset{frame=single,framexleftmargin=-10pt,framexrightmargin=-10pt,framesep=12pt,linewidth=0.98\textwidth,language=javascript}
\begin{lstlisting}[caption={Flexible usage of the \require function},label={lst:flexiblerequire}, basicstyle=\footnotesize]
1 try {
2   var util = require('util');
3  if (typeof util.inherits !== 'function') throw '';
4    module.exports = util.inherits;
5 } catch (e) {
6    module.exports = require('./inherits_browser.js');
7 }
\end{lstlisting}

\lstset{frame=single,framexleftmargin=-10pt,framexrightmargin=-10pt,framesep=12pt,linewidth=0.98\textwidth,language=javascript}
\begin{lstlisting}[caption={Static analysis fails to detect dependencies without executing the \texttt{require()} function},label={lst:require_in_runtime}, basicstyle=\footnotesize]
1 ...
2 return plugins.map(async pluginName => {
3    await server.register({
4       plugin: require(`../plugins/${pluginName}`),
5       options: {
6           ...(config[pluginName] || {})
7       },
8       routes: {
9           prefix: '/v4'
10       }
11   });
12 });
13 ...
\end{lstlisting}

\emph{(iv) \javascript's scripting nature allows \require\ calls to be generated dynamically as variable strings.}
Beyond the challenges created by \require\ being a function, the scripting nature of \javascript further complicates static analysis.
In \javascript, code can be treated as a string until it is executed, and that string can be implicitly or explicitly manipulated.
To illustrate this, we have devised the example in \autoref{lst:scriptingrequire}.
A reference to a dependency (e.g. using \require) in such code can only be identified at runtime, making it difficult for static analysis methods that rely on source code transformation to accurately detect dependencies \citep{koishybayev2020mininode, turcotte2021stubbifier}.

\bigskip

\lstset{frame=single,framexleftmargin=-10pt,framexrightmargin=-10pt,framesep=12pt,linewidth=0.98\textwidth,language=javascript}
\begin{lstlisting}[caption={Scripting as an obstacle to static dependency anlysis},basicstyle=\footnotesize, label={lst:scriptingrequire}]
1 function load(a, b){
2    eval(a + 'quire(' +b + ')');
3 }
4 load('re', 'some_package');
\end{lstlisting}

\subsection{State-of-the-art Technique for Debloating \nodejs Applications}
\label{subsec:stateofart}
As bloated dependencies lead to unnecessary maintenance overhead, detecting and removing them becomes essential for effective dependency management.
Given that developers can only remove entire dependencies rather than specific modules or functions to reduce maintenance costs, our trace-based analysis focuses on debloating entire bloated dependencies.
In this subsection, we introduce two state-of-the-art techniques for debloating \nodejs applications.
One is a widely-used static tool designed to identify unused direct dependencies, while the other employs dynamic analysis to debloat \nodejs applications.

% depcheck
\depCheck \footnote{\url{https://github.com/depcheck/depcheck}} is a widely-used tool that can detect direct bloated dependencies, which conducts static analysis of package source code and dependency declarations.
By parsing through \javascript modules and inspecting import statements, \depCheck\ identifies dependencies that are declared in \package file but not actually used within the source code.
Despite its popularity in industry and research areas \citep{jafari2021dependency}, \depCheck\ overlooks dependencies imported by the \require\ function if they are dynamically generated during runtime.
For example, consider line 4 of \autoref{lst:require_in_runtime}, which demonstrates how the \require\ function imports a dependency according to different parameters.
In such cases, \depCheck\ ignores the \require\ function, leading to inaccuracies in predicting imported dependencies.
This leads \depCheck\ to incorrectly report the unused dependencies, which is consistent with previous studies \citep{kabir2022developers}.
Our approach, on the other hand, dynamically tracks dependency imports during runtime, ensuring a more accurate detection of dependencies, including those generated dynamically.

Additionally, dynamic analysis techniques that focus on runtime behavior are relatively rare when it comes to addressing bloated dependencies.
Turcotte and colleagues introduce coverage-based analysis in the tool \stubbifier, which offers configurable dynamic analysis for debloating \nodejs applications \citep{turcotte2021stubbifier}.
This approach identifies and removes unused code, relying on the widely used test coverage tool, \texttt{nyc}\footnote{\url{https://github.com/istanbuljs/nyc}}, which instruments source code before test execution and records runtime behaviors.
By considering a module unused if none of its functions are invoked, \stubbifier generates a list of such modules, allowing for the identification of dependency use.
While this method improves upon static analysis, particularly in identifying dynamically imported dependencies via the \require\ function, its reliance on source code instrumentation may pose challenges when analyzing scripting languages like \javascript.

%%=============================================================%%
\section{Automatic Analysis of Bloated Dependencies in \commonjs Packages}\label{sec:test-debloating}

This section presents an overview of our approach for automatically detecting and removing bloated dependencies in \commonjs packages, implemented using our tool, \depPrune.
Before delving into the details, we introduce two foundational concepts.

\begin{definition}[Unaccessed Dependency]
\edit{A dependency is classified as unaccessed if none of its modules are accessed during the execution of the package.}
\end{definition}

\edit{It is important to note that the classification of a dependency as \emph{unaccessed} depends on the type of analysis used.
For example, unaccessed dependencies can be detected by instrumenting individual functions within a dependency and monitoring whether those functions are invoked at runtime \citep{vazquez2019slimming}.
Alternatively, they can be determined by analyzing execution reports, such as coverage reports, to identify dependencies that are not exercised during runtime \citep{turcotte2021stubbifier}.
In this work, we develop a technique based on monitoring the file system to determine whether a dependency is accessed or not.}

\begin{definition}[Bloated Dependency]
\edit{An unaccessed dependency is considered bloated if its complete removal from the dependency tree does not impede the successful execution of the package.}
\end{definition}

\edit{We note here, that the key difference between a \emph{bloated} and an \emph{unaccessed} dependency is that a dependency can only be confirmed as bloated if it is uninstalled from the package and the package continues to function correctly after re-execution.
Due to the highly dynamic and asynchronous nature of \javascript, it is challenging for program analysis techniques to precisely and completely capture the dependencies accessed at runtime. 
For example, if the analysis relies on test coverage data (\Cref{subsec:stateofart}), asynchronous functions that execute after the coverage report is generated may be missed.
As a result, the analysis may incorrectly classify a required dependency as unaccessed.
In such cases, removing the misclassified dependency can break the program, indicating that it is not bloated despite being labeled unaccessed.
}

\subsection{Overview}
\begin{figure}
  \centering
  \includegraphics[width=1\textwidth]{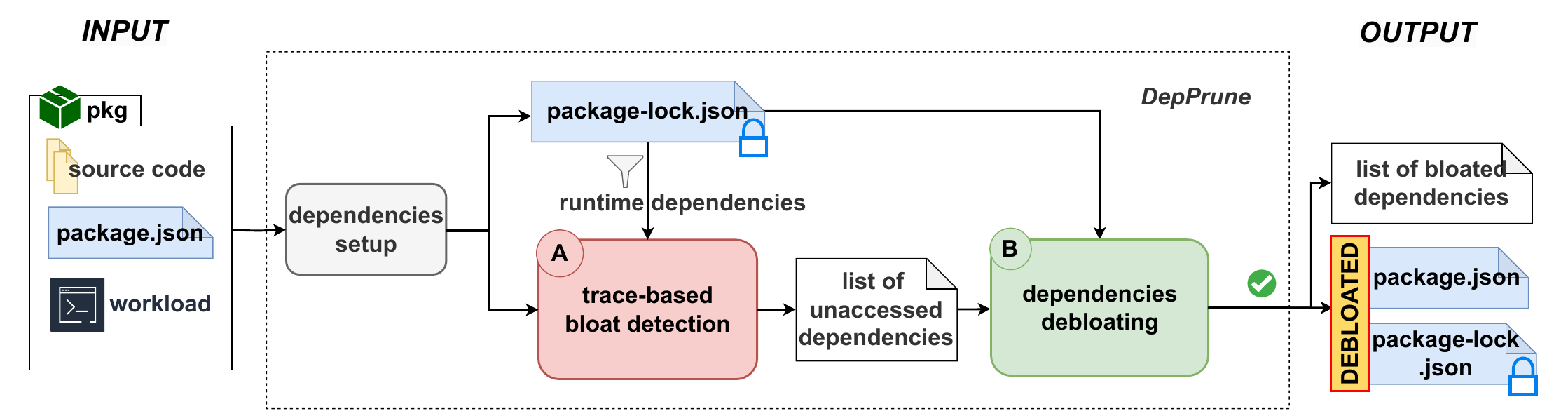}
  \caption{
  Overview of our approach for automatically debloating runtime dependencies in \commonjs packages.
  The input consists of the original package, including its source code, the \package file (listing direct dependencies only), and the workload of the package.
  The output includes a list of identified bloated dependencies, a debloated \package file with a reduced set of direct dependencies, and a debloated \packagelock file documenting all remaining dependencies, both direct and indirect.
  }
  \label{fig:overview}
\end{figure}

\autoref{fig:overview} outlines the workflow of our approach for automatically analyzing bloated dependencies.
Our approach begins with the original package, which includes its source modules, the \package file, and a workload that triggers the execution of the package.
We then install all dependencies using the official \texttt{npm install} command, which generates the \packagelock file.
This file provides detailed metadata about the installed dependencies, including their exact versions and intended usage (runtime or development).
By parsing the \packagelock file, we extract a list of runtime dependencies — dependencies that directly influence the package's behavior during production.
In our study, we focus exclusively on runtime dependencies and disregard development dependencies.

After the initial setup, we proceed to the core of our approach: the trace-based debloating technique, which consists of two key steps.
Step A: Tracing and Detecting Unaccessed Dependencies.
In this step, we execute the package using the input workload and perform a trace-based analysis to monitor which dependencies are accessed by the OS file system during execution.
Dependencies that are not accessed are identified by excluding the accessed ones from the original set of runtime dependencies.
Step B: Debloating Dependencies.
We transform the \packagelock file by removing the declaration of the unaccessed dependencies, effectively creating a debloated dependency tree.
To validate these changes, we rebuild the package using the updated file and re-execute it with the same workload to confirm that its functionality remains intact.

If the package runs successfully, this step yields the final output, which includes a list of bloated dependencies, the debloated \package file, and the debloated \packagelock file.
It is important to note that throughout this process, we do not modify the package's original code in any way.

\subsection{Trace-based Bloat Detection}
\label{subsec:runtime_identification}
Unlike many other programming languages that execute compiled code, \javascript runs directly as source files via the \javascript engine, using Just-In-Time (JIT) compilation.
In \commonjs, source code modules (i.e., files) are loaded and executed at runtime without any intermediate compilation step.
This runtime nature of \javascript allows us to monitor file accesses during execution, enabling the identification of modules that are accessed and, consequently,  to detect dependencies where no modules within are accessed.

To achieve this, we employ a trace-based dynamic analysis, as detailed in the following steps. 

\emph{1.Execution and tracing} The process begins by executing the package using the input workload.
During execution, our approach monitors all interactions between the OS file system and the running processes.
This monitoring generates a detailed log of all file operations, including file paths across various formats, as illustrated in \autoref{lst:rec-executions}.
Only modules actively accessed by the OS file system during execution are captured, while those not involved in the execution are excluded.

\emph{2.Filtering accessed modules} We then filter the file paths to isolate those relevant to dependencies.
Specifically, we target modules located in the dependency folder (\texttt{package/node\_modules}) with extensions \texttt{.js} or \texttt{.json}.
This step generates a refined list of accessed modules, represented by path.
For example, line $4$ of \autoref{lst:rec-executions} illustrates that the module \texttt{airtap/node\_modules\\/airtap-default/index.js} is accessed during execution.

\begin{lstlisting}[caption={Monitored interactions between OS and running files within \texttt{airtap}},
  label={lst:rec-executions},
  breaklines=true,
  basicstyle=\scriptsize,
  breakatwhitespace=true]
1 644728 openat(AT_FDCWD, "/lib/x86_64-linux-gnu/libc.so.6", O_RDONLY|O_CLOEXEC) = 3
2 644728 openat(AT_FDCWD, "/sys/fs/cgroup/memory/memory.limit_in_bytes", O_RDONLY|O_CLOEXEC) = 17
3 644728 openat(AT_FDCWD, "/proc/meminfo", O_RDONLY|O_CLOEXEC) = 17
4 644728 openat(AT_FDCWD, %*\fboxsep 1pt \colorbox{yellow!20}{"/disk/airtap/node\_modules/airtap-default/index.js"}*), O_RDONLY|O_CLOEXEC) = 17 
...
5 644746 openat(AT_FDCWD, %*\fboxsep 1pt \colorbox{yellow!20}{"/disk/airtap/node\_modules/mime-db/db.json"}*), O_RDONLY|O_CLOEXEC) = 23
6 644728 openat(AT_FDCWD, %*\fboxsep 1pt \colorbox{yellow!20}{"/disk/airtap/node\_modules/browserify/node\_modules}*)
%*\fboxsep 1pt \colorbox{yellow!20}{/readable-stream/readable.js"}*), O_RDONLY|O_CLOEXEC) = 17
...
\end{lstlisting}

\emph{3.Mapping accessed modules to dependencies} Each accessed file path is then analyzed to determine its corresponding dependency.
This step leverages the \npm dependency installation mechanism, where each dependency is installed in a folder named after the dependency itself, as explained in \Cref{subsec:npm-structure}.
For example, line $1$ of \autoref{lst:rec-files} illustrates that the module \texttt{airtap/node\_modules/airtap-default/index.js} maps to the dependency \texttt{airtap-default}.
At this stage, we generate a list of accessed dependencies, capturing all dependencies with at least one module accessed during runtime.

\begin{lstlisting}[caption={List of accessed dependency modules within \texttt{airtap}},label={lst:rec-files},breaklines=true, basicstyle=\scriptsize, breakatwhitespace=true]
1 airtap/%*\fboxsep 1pt \colorbox{yellow!20}{node\_modules/airtap-default}*)/index.js
2 airtap/%*\fboxsep 1pt \colorbox{yellow!20}{node\_modules/mime-db}*)/db.json 
3 airtap/%*\fboxsep 1pt \colorbox{yellow!20}{node\_modules/browserify/node\_modules/readable-stream}*)
  /readable.js
...
\end{lstlisting}

\emph{4.Identifying unaccessed dependencies} Finally, our approach identifies unaccessed dependencies by comparing the complete set of runtime dependencies, with the list of accessed dependencies generated in the previous step.
Dependencies that are not found in the accessed list are classified as unaccessed and considered candidates for debloating in the next phase.

\subsection{Dependencies Debloating}\label{sec:debloating}

The debloating phase focuses on pruning the dependency tree by removing unnecessary dependencies while ensuring the package retains its functionality.
Our approach employs two strategies: \textbf{direct-only debloating}, which targets bloated direct dependencies, and \textbf{full-scale debloating}, which addresses both direct and indirect bloated dependencies.
These strategies provide flexibility in managing dependency bloat, allowing developers to choose the most suitable approach for their needs.
We now detail this phase in the following steps:

\emph{1.Removing unaccessed dependencies from metadata files} This process begins by setting the initial set of bloated dependencies as the unaccessed dependencies identified in the trace-based analysis. 
These dependencies are located within the metadata files: the \package file, which specifies direct dependencies, and the \packagelock file, which records the entire dependency tree.
For direct-only debloating, we modify the \package file to remove declarations of direct unaccessed dependencies, ensuring they are no longer listed. 
This helps streamline the management of direct dependencies.
For full-scale debloating, we modify the \packagelock file to eliminate occurrences of both direct and indirect unaccessed dependencies.
In both strategies, the removal of dependencies is conducted all at once,  pruning the entire dependency tree in one step.

\emph{2.Rebuilding the package} Once the metadata files are updated, we use them to rebuild the package and verify the changes.
In direct-only debloating, the rebuild uses the updated \package file.
This ensures that direct unaccessed dependencies are removed from their declaration file and excluded from future builds.
In full-scale debloating, the rebuild uses the modified \packagelock file.
This prevents the installation of any unaccessed dependencies, including indirect ones, effectively pruning the dependency tree.

\emph{3.Validating functionality} After rebuilding, we execute the same inputted workload to verify that the package remains functional.
This step ensures that the removal of dependencies does not compromise the behavior of the package.
If functionality issues arise during validation, we restore the mistakenly removed dependencies and adjust the list of bloated dependencies to exclude them. 
The validation process is repeated until the debloating preserves the package's original functionality.

Once the process identifies the minimal set of dependencies that can be safely removed without impacting functionality, our approach compiles a final list of bloated dependencies and generates a debloated \package file (where direct bloated dependencies are removed) and a debloated \packagelock file (where all bloated dependencies are excluded).
Each file can ensure a successful build and execution of the package.

The debloated \package file simplifies managing direct dependencies and prevents their reinstallation during future builds, while the debloated \packagelock file ensures unnecessary dependencies are excluded, reducing maintenance overhead and potential security risks.
Together, these files support successful package builds and maintain consistency in dependency management.

\subsection{Implementation}
Our automatic analysis focuses on the detection and removal of runtime-only dependencies in \commonjs packages, implemented with a novel tool \depPrune.
As described in \autoref{fig:overview}, the two main processes of \depPrune's pipeline 
are integrated with two separate Shell scripts.
The subtasks of each process are automated in Python or Shell scripts, which are available on Github: \url{https://github.com/ASSERT-KTH/DepPrune}.

To install all dependencies for a package, we use the official command in \npm, which is \texttt{npm install}.
To identify runtime dependencies, we implement a Python script to analyze \packagelock, where dependencies are scoped for use in runtime or development.

During the trace-based detection phase, we monitor the interactions between the file system and the package execution.
We rely on the  \texttt{strace} Linux utility, which is widely used for debugging and troubleshooting tasks within the OS file system.
\texttt{strace} can monitor OS file system calls that are made during a specific task.
By using \texttt{strace}, \depPrune tracks which modules are called by the OS file system during execution.

During the debloating phase, we ensure that dependency resolution remains reproducible by copying the original package source code and the generated \packagelock, then using the official \npm command \texttt{npm ci} to strictly install each dependency exactly as specified in the \packagelock file \citep{npmcilockfile}.
To produce debloated versions of \package and \packagelock, we leverage the Python built-in \texttt{json} library to parse the original files, identify bloated dependencies, and remove each occurrence of these dependencies from the \texttt{dependencies} of other packages.
This involves removing direct bloated dependencies from the \package, and both direct and indirect bloated dependencies from \packagelock.

We conduct an extensive assessment of \depPrune to ensure its correctness, incorporating both manual and automated testing. 
First, we run \depPrune on a dataset of small, popular \npm packages published in prior work \citep{turcotte2021stubbifier}. 
After \depPrune successfully processes each package, we manually verify that dependencies identified as bloated can indeed be removed without affecting the functionality of the package. 
Second, to validate the transformation of removing bloated dependencies from 
\package and \packagelock, we create a testing script that compares the JSON files before and after the transformation, with bloated dependencies as input.
Third, we run the official \npm command \texttt{npm ci} to ensure the transformed JSON files work without errors, and use the command \texttt{npm ls} to list the remaining dependencies, confirming that the removed dependencies are no longer present.
Finally, we have automated tests that check the functionality of our script, such as reading accessed dependencies from the execution log and computing unaccessed dependencies.
These validations ensure that \depPrune performs as intended, and provide a reliable foundation for the experiments presented in this study.

%%=============================================================%%

%%=============================================================%%
\section{Methodology}
\label{sec:ptcls}
This section outlines our evaluation methodology.
We motivate the research questions, then we present the process to curate a dataset and eventually, we introduce the protocol we follow to answer each research question. 

To perform our trace-based analysis, we choose the execution of package unit tests as our workload for analyzing dependency bloat, inspired by prior studies on analyzing code bloat through testing \citep{bruce2020jshrink, xin2022studying, soto2023coverage}.
This guarantees that the package expected behavior is preserved after removing the dependencies that are detected as bloated.

\subsection{Research Questions}
\subsubsection{RQ1: \RQone}

RQ1 investigates the prevalence of bloated dependencies, both direct and indirect, identified by our trace-based dynamic analysis approach.
As no prior studies have examined the prevalence of bloated dependencies within \npm packages, RQ1 aims to fill that gap by providing an understanding of how common these dependencies are.
Additionally, we check whether the proportion of bloated dependencies correlates with the size of the dependency tree.
This will indicate whether dependency bloat is primarily a scale-related issue in larger dependency trees or a widespread challenge affecting packages regardless of their dependency tree size.

\subsubsection{RQ2: \RQtwo}
RQ2 examines the impact of debloating direct dependencies on the overall dependency tree of a package.
Direct dependencies are explicitly managed by developers through the \package file, making their removal practical and a relevant part of dependency maintenance. Meanwhile, there is no native support in \npm to properly remove and maintain indirect dependencies.

Some indirect dependencies can be exclusively associated with direct bloated dependencies or shared between bloated and non-bloated direct dependencies. As part of this research question, we quantify how many indirect dependencies actually disappear from the dependency, when removing only direct bloated dependencies.

\subsubsection{RQ3: \RQthree}

In RQ3, we compare our trace-based analysis approach with two state-of-the-art approaches for identifying bloated dependencies in \commonjs packages: a dynamic coverage-based analysis technique and a static analysis technique (introduced in \Cref{subsec:stateofart}).
This research question aims to evaluate the specific strengths and limitations of each approach, focusing on their accuracy in identifying bloated dependencies.
The comparison between trace-based and coverage-based dynamic analyses aims to uncover insights that can refine dynamic dependency analysis, particularly in the inherently dynamic \commonjs environment.
On the other hand, the comparison between trace-based and static analyses focuses on understanding the limitations of static methods in capturing the dynamic behaviors of \commonjs modules.
Through these comparisons, RQ3 aims to highlight the unique advantages of trace-based analysis in dynamic contexts while offering insights into the challenges faced by static methods when applied to \commonjs packages.
We hope these findings inform future work on improving dependency analysis techniques for dynamic environments.

\subsection{Dataset Collection}
\label{sec:data}
To answer our research questions, we curate a dataset of real-world \commonjs packages, serving as the foundation for evaluating the effectiveness of our trace-based dynamic analysis using \depPrune.
Since our evaluation relies on executing the packages with their unit tests, we select packages with high test coverage.
This ensures that the execution adequately exercises the package, allowing us to confidently identify bloated dependencies based on runtime behavior.

We begin by collecting \npm packages that meet the following criteria: (i) use the \commonjs module system, (ii) have both direct and indirect dependencies, (iii) build successfully, and (iv) have high test coverage.
As a starting point, we rely on \texttt{deps.dev}\footnote{\url{https://docs.deps.dev/}}, an online service launched by Google in 2021 to help developers understand the network of dependencies within their projects.
The dataset collection through \texttt{deps.dev} is conducted in the study by \citet{rahman2024characterizing}, which aligns with the focus of our study.

Initially, we gather the top 100,000 \npm packages with the highest number of dependencies as of May 29, 2023.
We then exclude packages without a corresponding GitHub repository and scoped packages (those starting with "@"), as some scoped packages are part of larger modules and may not have standalone test suites.
After these exclusions, we are left with 35,321 \npm packages, which we further filter based on specific criteria described below.

\begin{enumerate}
    \item Valid GitHub Link: We ensure all packages have valid GitHub links, reducing our set to 27,353 accessible packages.
    \item Test Suite: We keep only packages with a \package file and a \texttt{test}\footnote{\href{https://docs.npmjs.com/cli/v10/commands/npm-test}{https://docs.npmjs.com/npm-test}} script, resulting in 15,913 packages that contain a test suite.
    \item \commonjs Module System: We examine the entry files of these packages to confirm their use of the \commonjs module system, leaving us with 5,836 packages.
    \item Test Coverage: We focus on packages with a "test coverage" badge on GitHub and filter for high coverage (\(\geq 70\%\)). 
    After testing with \texttt{nyc}\citep{nycgitaddress}, 733 packages meet our criteria. 
    \item Successful Build: We retain only packages that successfully build and pass all their tests. This results in \nbPackages packages that form our dataset.
\end{enumerate}

To the best of our knowledge, our dataset is the first curated dataset of well-tested \npm packages that use the \commonjs module system.
We make this dataset publicly available on Zenodo: \href{https://zenodo.org/doi/10.5281/zenodo.15090140}{10.5281/zenodo.15090140}.

\begin{figure}
    \centering
    \includegraphics[width=1\textwidth]{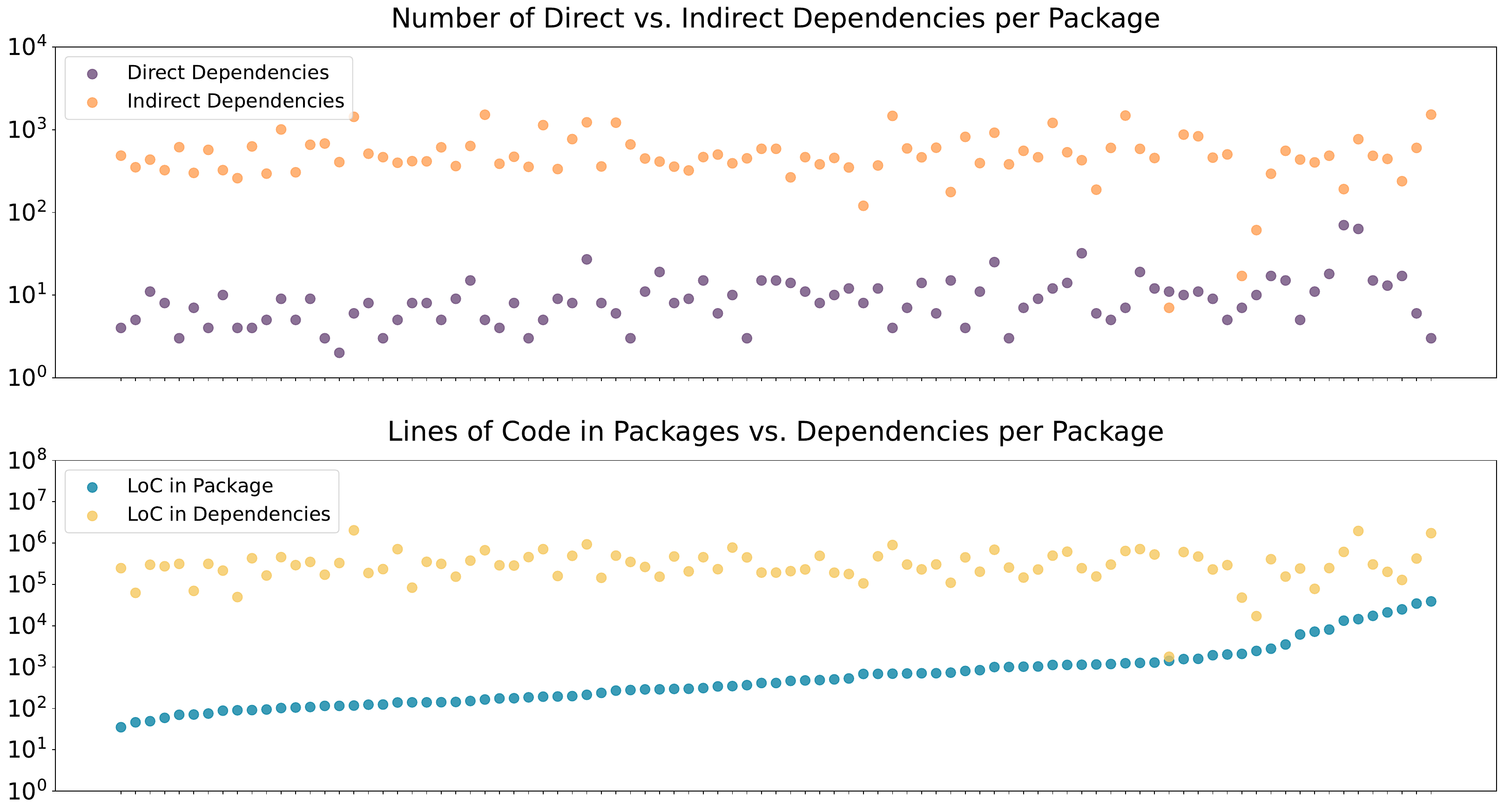}
    \caption{\edit{
    The number of direct (purple dots) and indirect dependencies (orange dots) in \nbPackages packages, along with lines of code (LoC) in packages (blue dots) and their dependencies (yellow dots), ordered by LoC in package in ascending order.
    The underlying data for this figure is publicly available: \href{https://zenodo.org/doi/10.5281/zenodo.15090140}{10.5281/zenodo.15090140}
    }}
    \label{fig:pkgScatterPlot}
\end{figure}

\autoref{fig:pkgScatterPlot} provides an overview of our dataset, in which packages are ordered by lines of code (LoC), arranged in ascending order.
The upper part illustrates the number of direct and indirect runtime dependencies for each package.
The number of indirect dependencies is counted based on unique installations, meaning that we count each instance wherever a dependency is installed multiple times.
% Additionally, as explained in \Cref{sec:test-debloating}, our analysis focuses exclusively on runtime dependencies, rather than development dependencies.
Across our dataset of \nbPackages packages, there are a total of \totalNbDeps runtime dependencies, including \totalNbDirectDeps direct and \totalNbIndirectDeps indirect ones.
The number of direct dependencies ranges from 1 (in \href{https://github.com/kuzzleio/protocol-mqtt-proxy/tree/99b25d17b0ba277625d85c47b1b3513448b605d8}{\texttt{kuzzle-plugin-mqtt}}) to 69 (in \href{https://github.com/cedric05/cli/tree/54079eadb5ba8f4c7e45c2876ea589e5e4844342}{\texttt{cedric05-cli}}), with a median of 7, while the number of indirect dependencies ranges from 6 (in \href{https://github.com/caviarjs/caviar/tree/f9ee9d3af61f6a3c5c779da176e4d0fff1b8a866}{\texttt{caviar}}) to 1,521 (in \href{https://github.com/easy-team/easywebpack-js/tree/f45f5f03f486e532d1f7b1e480dc374cb567e37a}{\texttt{easywebpack-js}}), with a median of 462.
Most dependencies are not declared directly by developers but are instead pulled in indirectly through other packages, which aligns with previous studies \citep{soto2021comprehensive}.

In the lower part of \autoref{fig:pkgScatterPlot}, we present the lines of code (LoC) \edit{in .js or .ts files} in each package and the lines of code in their dependencies.
The LoC count is obtained using \texttt{cloc} \citep{clocgitaddress} and excludes \javascript code comments.
In total, the runtime dependencies across the \nbPackages packages include more than 34 million lines of code.
The LoC in packages ranges from 34 (in \href{https://github.com/rooseveltframework/mkroosevelt/tree/477df2d85a2d128e41bbdb83edec4789238e10b9}{\texttt{mkroosevelt}}) to 38,716 (in \href{https://github.com/hubcarl/easywebpack-js/tree/f45f5f03f486e532d1f7b1e480dc374cb567e37a}{\texttt{easywebpack-js}}), with a median of 410.
The minimum number of LoC in the runtime dependencies is 1,771 (in \href{https://github.com/caviarjs/caviar/tree/f9ee9d3af61f6a3c5c779da176e4d0fff1b8a866}{\texttt{caviar}}), while the maximum is more than 2 million (in \href{https://github.com/easy-team/easywebpack-vue/tree/b77f5f80615b353279da2d9183b6209b321bda18}{\texttt{easywebpack-vue}}).
This observation is consistent with previous studies indicating that most of the code in a package comes from its dependencies rather than the original source code \citep{koishybayev2020mininode}.

From \autoref{fig:pkgScatterPlot}, we see that neither the number of direct dependencies, the number of indirect dependencies, nor the LoC count in the dependencies are correlated with the package size.
Notably, our dataset reveals that smaller packages are highly likely to introduce a significant number of dependencies, consistent with previous findings \citep{abdalkareem2020impact}. 
For instance, the package \href{https://github.com/vigour-io/brisky/tree/26113936272fc0e51f896dede83a006d3739fbb6}{\texttt{brisky}} contains only 48 lines of its own source code, yet it imports 10 direct dependencies that lead to 433 indirect dependencies, resulting in a total of 298,840 lines of code from all dependencies.
\edit{This underscores how small packages can accumulate extensive dependency trees, highlighting the potential for unnecessary dependencies that may not contribute to the package's functionality.
Project size alone is therefore not a reliable indicator of project complexity.}

Although our minimum threshold for including a package in the dataset is 70\% test coverage, the median coverage is \testCov.
\edit{ Our dataset includes 32 packages with 100\% coverage, 36 packages with 90–99.9\% coverage, 15 packages with 80–89.9\% coverage, and 8 packages with 70–79.9\% coverage.}
This high level of test coverage makes our dataset well-suited for our evaluation, ensuring the relevance of our findings.

Our dataset also spans a diverse range of development activities and popularity levels among software packages.
The number of commits ranges from 6 to 12,477, with a median of 108 commits per package.
In terms of popularity, as measured by GitHub stars, the packages range from 0 to 32,950 stars, with a median of 3 stars, reflecting a wide distribution of popularity within our dataset.

\subsection{Protocol for RQ1}
\label{subsec:ptcl1}
To address RQ1, we run \depPrune on each package in our dataset, using the full-scale debloating strategy, to identify and quantify all bloated dependencies that can be removed from the package dependency tree.
In answer to this question, we report the number of direct and indirect bloated dependencies, as well as their percentage, across the \nbPackages packages in our dataset, providing an overview of the prevalence of dependency bloat.

Note that dependencies installed multiple times within the tree are counted multiple times.
As explained in \Cref{subsec:npm-structure}, dependencies are often installed multiple times due to version mismatches or required by different parent packages.
This counting approach reflects the real-world complexity and challenges of managing multiple instances of the same dependency.
We also account for each installed instance to measure the dependency tree size, offering a precise representation of the dependency tree's scale and its implications, such as build time and maintainability.
This approach is also consistent with the dependency tree structure documented in \packagelock, where each instance of a dependency installed multiple times is treated as a distinct dependency.

\subsection{Protocol for RQ2}
\label{subsec:ptcl2}
To address RQ2, we run \depPrune on each package in our dataset, using the direct-only debloating strategy, with a focus on packages that have direct bloated dependencies.
To quantify the impact of debloating direct dependencies, we calculate the reduction in the number of installed dependencies using the ratio $\text{R}_{\text{d}} = \frac{\text{$Dep\_r$}}{\text{$Dep\_o$}}$.
Here, $Dep\_r$ represents the total number of dependencies removed, including both direct bloated dependencies and their associated indirect dependencies.
$Dep\_o$ represents the original total number of installed dependencies.
This ratio clearly measures the impact of debloating direct dependencies on the dependency tree, illustrating the follow-on effect where the removal of direct dependencies also eliminates their associated indirect dependencies.

\subsection{Protocol for RQ3}
\label{subsec:ptcl3}
To address RQ3, we conduct a comparative study to evaluate the effectiveness of trace-based analysis with two state-of-the-art approaches for detecting bloated dependencies: the comparison with coverage-based dynamic analysis (\citep{Stubbifiergitaddress}) and the comparison with static analysis (\citep{depcheckgitaddress}).
To validate the effectiveness of detection, we use the package's test suite.
If the removal of a detected dependency ensures all tests pass successfully, the detection is deemed effective.

\emph{Comparison with coverage-based dynamic analysis.} 
For this comparison, we use an adapted version of \stubbifier\footnote{\href{https://github.com/ASSERT-KTH/stubbifier}{ASSERT-KTH/Stubbifier}} to perform coverage-based dynamic analysis.
This adaptation includes two logging instructions to capture all modules detected as bloated by \stubbifier.
The protocol involves the following steps: 1) We execute the adapted \stubbifier on each package in our dataset and produce logs, capturing the identified bloated modules. 
2) We identify dependencies that consist entirely of bloated modules and classify them as bloated dependencies. 
3) We apply the full-scale debloating strategy (\Cref{sec:debloating}) to remove the identified bloated dependencies. 
4) We execute the test suite of the package to validate the removal. 
5) We record the total number of identified bloated dependencies by \stubbifier, including both direct and indirect ones. 
6) We analyze the differences in the dependencies identified as bloated by \stubbifier and \depPrune.

\emph{Comparison with static analysis}
For the comparison with static analysis, we use \depCheck, a tool that identifies only direct dependencies. 
This comparison focuses on direct dependencies and involves the following steps: 1) We execute \depCheck on each package in our dataset using its official command \citep{depcheckgitaddress}, to identify direct bloated dependencies.
2) We apply the direct-only debloating strategy (\Cref{sec:debloating}) to remove the identified direct bloated dependencies.
3) We execute the test suite of the package to validate the removal.
4) We record the number of direct bloated dependencies identified by \depCheck.
5) We analyze the differences in the direct dependencies identified as bloated by \depCheck and \depPrune.

%%=============================================================%%

%%=============================================================%%
\section{Experimental Results}\label{sec:result}
This section evaluates our approach with \depPrune, in debloating \commonjs dependencies within a dataset of \nbPackages real-world, well-tested \commonjs packages.
We now present the results for our research questions, following the experimental protocol outlined in \Cref{sec:ptcls}.
We conclude this section with a discussion about the runtime overhead of our approach.

\subsection{RQ1: \RQone}
To answer RQ1, we execute \depPrune across the \nbPackages packages in our dataset, containing \totalTestCases test cases and \totalNbDeps dependencies (direct and indirect). 
Our trace-based analysis determines that \totalAccessedModules modules (files), which belong to \totalNbNonBloatedDeps dependencies, are accessed at runtime.
Conversely, \totalNbBloatedDeps dependencies comprise only modules that are never accessed at runtime and are therefore identified as unaccessed dependencies, as explained in \Cref{sec:debloating}.
\edit{Overall, \depPrune detects that \ratioBloatedDeps (\totalNbBloatedDeps / \totalNbDeps) of all dependencies in our dataset are bloated, with a 100\% of the test cases in all packages still passing after the their removal}.

% overall: 91 packages, direct, indirect
Out of \nbPackages packages, \nbPackagesBloated (98.9\%) contain at least one bloated dependency. 
The number of bloated dependencies ranges from 2 (in \href{https://github.com/caviarjs/caviar/tree/f9ee9d3af61f6a3c5c779da176e4d0fff1b8a866}{\texttt{caviar}}) to 1,470 (in \href{https://github.com/jclo/umdlib/tree/572af0377f2194339eab79554c6b7e0bb7160ffe}{\texttt{umdlib}}), with a median value of 180.
Only 1 package (\href{https://github.com/cedric05/cli/tree/54079eadb5ba8f4c7e45c2876ea589e5e4844342}{cedric05-cli}) stands out for having no bloated dependencies.
Our findings reveal the widespread prevalence of bloated dependencies in the \commonjs packages.
Among the \totalNbDirectDeps direct dependencies analyzed, \ratioDirectBloatedDeps (\totalNbBloatedDirectDeps/\totalNbDirectDeps) are bloated, whereas among the \totalNbIndirectDeps indirect dependencies, we find that more than half (\ratioIndirectBloatedDeps) of them are bloated.
% indication
Our quantitative study reveals that indirect dependencies are predominant, accounting for 99.5\% of all bloated dependencies in the \commonjs packages in our dataset.

We further analyze the subset of 32 packages with 100\% test coverage to assess whether similar results hold for all packages.
Within this subset, \subsettotalNbDeps dependencies are analyzed, of which \subsettotalNbBloatedDeps (\subsetratioBloatedDeps) are identified as bloated.
Among the \subsettotalNbDirectDeps direct dependencies, \subsettotalNbBloatedDirectDeps (\subsetratioDirectBloatedDeps) are bloated, while among the \subsettotalNbIndirectDeps
indirect dependencies, more than half (\subsetratioIndirectBloatedDeps) are bloated.
Packages with 100\% test coverage have a slightly lower ratio of direct bloated dependencies compared to the overall dataset but exhibit a higher ratio of indirect bloated dependencies.
These results align with the overall dataset, where indirect dependencies account for the majority of bloated dependencies.
We see that even packages with 100\% test coverage are not immune to dependency bloat, emphasizing that testing alone cannot guarantee the overall quality of a package with respect to bloat.
Our trace-based analysis also stresses the importance of a comprehensive workload for effectively detecting and reducing dependency bloat.

Additionally, we investigate the correlation between the percentage of bloated dependencies in the packages and their dependency tree size.
To measure this, we calculate the \texttt{Spearman rank correlation coefficient} ($r_s$), applied in prior study \citep{wittern2016look}.
Our results show that the value of $r_s$ is 0.14739, with a two-tailed p-value of 0.16325.
This suggests a weak positive correlation that is not statistically significant.
Thus, we conclude that the size of the dependency tree does not significantly impact the proportion of bloated dependencies, indicating that packages with different dependency tree sizes are equally likely to suffer from dependency bloat.

\bigskip
% packages with direct bloated
Out of \nbPackages packages, \nbPackagesDirectBloated (50.5\%) contain at least one direct bloated dependency.
The number of direct bloated dependencies across the \nbPackagesDirectBloated packages has a median value of 2, with most packages having a number of direct bloated dependencies between 1 and 3; yet, there are cases where packages have a significant number of direct bloated dependencies.
The package \href{https://github.com/screwdriver-cd/screwdriver/tree/cc915df162b3aa8117ad8f2294ced64c7f75513d}{\texttt{screwdriver}} has 19 direct bloated dependencies, the maximum in our dataset.
The existence of packages with direct bloated dependencies is consistent with previous studies \citep{McIntoshicse2014, soto2021comprehensive}, and emphasizes the need for effective tools to increase the quality of \package files.

To further understand the nature of direct bloated dependencies, we manually analyze the \totalNbBloatedDirectDeps direct bloated dependencies across \nbPackagesDirectBloated packages.
This analysis involves reviewing the official documentation, source code, and usage examples of each dependency to determine the intended purpose.

Our findings reveal specific reasons for their identification as bloated during testing runtime.
Notably, 41.7\% (50/120) of the direct dependencies are explicitly documented as development dependencies, but are misused as runtime dependencies in the \package file. 
For example, the dependency \texttt{webpack} of the package \href{https://github.com/moneybutton/bsv-mnemonic/tree/6c73e2a47aba7aad6758aa820c9b59faa47ef0fc}{\texttt{bsv-mnemonic}} is used during the build process to bundle and compress source code into smaller size, a task that qualifies it as a development dependency.
Such misuse introduces unnecessary runtime dependencies that could be avoided.

We categorize these misused development dependencies by their intended purposes, as illustrated in \autoref{table:bloated_reason}.
Dependencies used in the transpiling process are the most prevalent, accounting for 56\% (28/50).
These dependencies primarily support tasks such as code compilation, minification, and optimization during the build phase. 
Notably, the testing-related dependencies reported in row 2 are not utilized, even during the testing workload. 
We see that dependencies used in the transpiling process are the most prevalent, accounting for 56\% (28/50). 
This observation highlights the need for developers to carefully classify dependencies based on their actual usage to reduce unnecessary runtime bloat and improve dependency management practices.

\begin{table}[t] %% placement specifier
\centering %% For center alignment of the tabular.
\scriptsize
\caption{Breakdown of the misused development dependencies identified as bloated.
}\label{table:bloated_reason}
\begin{tabularx}{\textwidth}{l|l|p{8cm}} %% Table column specifiers
\toprule
 STAGE & \# DEP & EXPLANATION  \\ 
\midrule
Transpiling & 28 & Dependencies used for tasks such as code compilation, minification, and optimization during the build phase. E.g., the dependency \href{https://github.com/webpack/webpack}{\texttt{webpack}} in the package \href{https://github.com/moneybutton/bsv-mnemonic/tree/6c73e2a47aba7aad6758aa820c9b59faa47ef0fc}{\texttt{bsv-mnemonic}}  \\ \hline
Testing & 9 & Dependencies used for testing, such as generating test coverage report. E.g., the dependency \href{https://github.com/yeoman/generator-mocha}{\texttt{generator-mocha}} in the package \href{https://github.com/albatrosary/generator-angular-eggs/tree/531395c4aa6945c75cf7f76c2a1567f19947de10}{\texttt{generator-angular-eggs}}  \\ \hline
Bootstrapping & 8 & Dependencies used for setup, scaffolding, or the development process. E.g., the dependency \href{https://github.com/yeoman/yeoman-welcome}{\texttt{yeoman-welcome}} in the package \href{https://github.com/react-webpack-generators/generator-react-webpack-alt/tree/8b233731dc899d88945372bc05d601715bb822d5}{\texttt{generator-react-webpack-alt}}  \\ \hline
Building Configuration & 2 & Dependencies used for preparing the system or environment for the build, such as setting environment variables. E.g., the dependency \href{https://github.com/kentcdodds/cross-env}{\texttt{cross-env}} in the package \href{https://github.com/easy-team/easywebpack-js/tree/f45f5f03f486e532d1f7b1e480dc374cb567e37a}{\texttt{easywebpack-js}}  \\ \hline
Deploying & 2 & Dependencies used for managing the deployment process. E.g., the dependency \href{https://github.com/eggjs/scripts}{\texttt{egg-scripts}} in the package \href{https://github.com/easy-team/res/tree/bf15ea88a78cae938fdaa4065cf8a055d69f33dc}{\texttt{res}}  \\ \hline
Linting & 1 & Dependencies used for detecting and fixing programming errors or stylistic issues. E.g., the dependency \href{https://www.npmjs.com/package/@babel/eslint-plugin}{\texttt{babel-eslint}} in the package \href{https://github.com/TestArmada/magellan-seleniumgrid-executor/tree/5c3d11ebfe575d9251fe41997435179c90f76ea8}{\texttt{magellan-seleniumgrid-executor}}  \\ \hline
\bottomrule
\end{tabularx}
\end{table}

For the remaining dependencies that are documented as being used as runtime, we find that 19.2\% (23/120) dependencies are \texttt{command-line tools}, implemented for tasks such as interactions with the operating systems or parsing user inputs.
However, these tasks are not related to the packages' core logic, and often fall outside the scope of unit tests.
As a result, such dependencies remain unexercised during test execution.
Our findings highlight the need for broadening test coverage to include scenarios beyond standard unit tests, ensuring that runtime dependency usage is thoroughly evaluated.

\begin{answerbox}
\textbf{Answer to RQ1:}
% overall packages
For the \nbPackages \commonjs packages under study, \depPrune identifies \ratioBloatedDeps (\totalNbBloatedDeps/\totalNbDeps) of their dependencies as bloated, which can be successfully removed.
Specifically, \ratioDirectBloatedDeps (\totalNbBloatedDirectDeps/\totalNbDirectDeps) of the direct dependencies are bloated while \ratioIndirectBloatedDeps (\totalNbBloatedIndirectDeps/\totalNbIndirectDeps) of the indirect dependencies are bloated.
Our analysis reveals no significant correlation between the percentage of bloated dependencies and the dependency tree size.
% Implication
The misuse of development dependencies as runtime dependencies is a key contributor to dependency bloat.

\end{answerbox}

\subsection{RQ2: \RQtwo}
% pratical
In the practical development of \npm packages, developers maintain the \package file, adding or updating direct dependencies and metadata about their packages.
This maintenance often overlooks opportunities to reduce code bloat by removing direct unused dependencies.
Moreover, indirect dependencies further contribute to bloat as they can be either exclusively associated with direct bloated dependencies or by being shared between bloated and non-bloated direct dependencies.
To address the uncertainty about the extent to which removing only direct bloated dependencies can reduce the size of the dependency tree,
RQ2 investigates the extent to which removing direct bloated dependencies can reduce the size of the dependency tree, applying the direct-only debloating strategy.

% overall
Per RQ1, \nbPackagesDirectBloated packages contain at least one direct bloated dependency.
Using the direct-only debloating strategy, we evaluate how the removal of direct bloated dependencies alone affects the dependency trees of these packages.
In total, removing \totalNbBloatedDirectDeps direct bloated dependencies from \nbPackagesDirectBloated packages leads to the additional removal of \totalNbBloatedInrectDepsByDirect indirect dependencies.
This cascading effect is detailed in the \texttt{\#BD$\rightarrow$I} column of \autoref{table:direct_bloated}, which quantifies the number of indirect dependencies removed due to each direct bloated dependency.
% sorted in descending order based on the $R_d$ value. 
\begin{table}
\centering
\captionsetup{font=scriptsize}
\caption{Of the \nbPackages packages in our dataset, \nbPackagesDirectBloated have direct bloated dependencies.
We rank the table by $R_d$ (the ratio of dependency tree size reduction).
'Cov \%' denotes test coverage.
'\#D' and '\#T' denote the number of originally installed direct dependencies and total dependencies, respectively.
'\#UD' is the number of direct dependencies identified as bloated by \href{https://github.com/depcheck/depcheck}{\depCheck}.
'\#BD' is the number of direct dependencies identified as bloated by \depPrune and '\#BD$\rightarrow$I' is the number of corresponding indirect dependencies resulting from these direct bloated dependencies.
'$Prune\_d$' is the number of dependencies pruned from the dependency tree, followed by the corresponding $R_d$ value.
}\label{table:direct_bloated}
\scriptsize
\begin{tabular*}{\textwidth}{@{\extracolsep\fill}p{0.1cm}p{2cm}lp{0.3cm}p{0.5cm}p{0.3cm}p{0.5cm}p{1.8cm}l}
\toprule%
& \multicolumn{2}{l}{\scriptsize Package } 
& \multicolumn{2}{l}{\scriptsize Original } 
% & \multicolumn{1}{l}{\scriptsize \depCheck }  
& \multicolumn{4}{l}{\scriptsize \depPrune } \\
% \cmidrule{2-3} \cmidrule{4-6}\cmidrule{7-8}
% & Name & Cov \% & \#D & \#I & \#T & \#D & \#I\footnotemark[1] & $RATIO_d$ \\
\cmidrule{2-3} \cmidrule{4-5}  \cmidrule{6-9}
& \multicolumn{1}{l}{\scriptsize Name}
& \multicolumn{1}{l}{\scriptsize Cov \%} 
& \multicolumn{1}{l}{\scriptsize \#D}
& \multicolumn{1}{l}{\scriptsize \#T}
% & \multicolumn{1}{l}{\scriptsize \#UD}
& \multicolumn{1}{l}{\scriptsize \#BD} 
& \multicolumn{1}{l}{\scriptsize \#BD$\rightarrow$I}
& \multicolumn{1}{l}{\scriptsize $Prune\_d$}
& \multicolumn{1}{l}{\scriptsize $R_d$ \%} \\
\midrule
1 & \href{https://github.com/jclo/umdlib/tree/572af0377f2194339eab79554c6b7e0bb7160ffe}{umdlib} & 100 & 3 & 1,470 & 3 & 1,467 & 1,470 / 1,470 & 100 \\
2 & \href{https://github.com/Gleider/podcast-search/tree/8a79ed1ed37c9d345a832e7e2dfef823e0e200c7}{podcast-search} & 100 & 2 & 681  & 1 & 679 & 680 / 681 & 99.85 \\
3 & \href{https://github.com/mediagoom/node-play/tree/c67b86ca81a191a98ae5573ce870deaf734f4347}{node-webplay} & 100 & 5  & 192  & 2 & 162 & 164 / 192 & 85.42 \\
% 4 & \href{https://github.com/bigdatr/elephas/tree/163dcd567e05f1ad98ee8fe4f29b216ffbbb610a}{elephas} & 91.09 & 21  & 173 & 1 & 14 & 131 & 145 / 173 & 83.82 \\
4 & \href{https://github.com/uncledu/tagd/tree/f7091af35fb397bf837153e08b1d60a4756a4995}{tagd} & 87.18 & 13  & 277  & 11 & 151 & 162 / 277 & 58.48 \\
5 & \href{https://github.com/vigour-io/brisky-base/tree/587e9efaf856d9084be976407d420c28295cbdeb}{vigour-base} & 99.78 & 6 & 22  & 1 & 11 & 12 / 22 & 54.55 \\
6 & \href{https://github.com/moneybutton/bsv-mnemonic/tree/6c73e2a47aba7aad6758aa820c9b59faa47ef0fc}{bsv-mnemonic} & 97.71 & 5  & 606 & 2 & 309 & 311 / 606 & 51.32 \\
7 & \href{https://github.com/codemodsquad/jscodeshift-build-import-list/tree/896cafd932f678b92ae7f4e62e1cc04fd47fbf84}{jsc-list} & 84.84 & 8 & 341 & 1 & 104 & 105 / 341 & 30.79 \\
8 & \href{https://github.com/gulpjs/gulp/tree/85896d4f099a80d58dd08d1f6d80586b07678995}{gulp} & 100 & 4  & 354  & 1 & 100 & 101 / 354 & 28.53 \\
9 & \href{https://github.com/urdubiometer/urdubiometerjs/tree/826f8e5698a1fa8bc1cbaee01d781ed7458a2a10}{urdubiometerjs} & 100 & 2  & 664  & 1 & 181 & 182 / 664 & 27.41 \\
10 & \href{https://github.com/screwdriver-cd/screwdriver/tree/cc915df162b3aa8117ad8f2294ced64c7f75513d}{screwdriver} & 96.21 & 62 & 828  & 19 & 187 & 206 / 828 & 24.88 \\
11 & \href{https://github.com/Malvid/Malvid/tree/32eca112788308cc0f4b84226787af67aa211b49}{Malvid} & 97.43 & 9  & 69  &2 & 11 & 13 / 69 & 18.84 \\
12 & \href{https://github.com/Pikamuchu/pika-generator-cartridge-lib-module/tree/348d1c080e990e92f23bc1e2740971d7867621d6}{gen-module} & 94.64 & 7 & 775  & 1 & 145 & 146 / 775 & 18.84 \\
13 & \href{https://github.com/vigour-io/brisky-props/tree/9a15bee1724fe970135c70458125764290a39527}{brisky-props} & 100 & 3  & 572  & 2 & 72 & 74 / 572 & 12.94 \\
14 & \href{https://github.com/caviarjs/caviar/tree/f9ee9d3af61f6a3c5c779da176e4d0fff1b8a866}{caviar} & 99.53 & 10  & 16  & 1 & 1 & 2 / 16 & 12.5 \\
15 & \href{https://github.com/shadowwzw/webpack-dev-server-fork/tree/c71cd329c7baae0c7e52d08956ca2e3c9dc8f293}{web-fork} & 70.42 & 17 & 500  & 7 & 51 & 58 / 500 & 11.6 \\
16 & \href{https://github.com/vigour-io/brisky-focus/tree/561db27535942e5186e79248b203bc0b190475c5}{brisky-focus} & 100 & 8 & 664  & 2 & 74 & 76 / 664 & 11.45 \\
17 & \href{https://github.com/vigour-io/brisky-events/tree/c181c18777c77b32dc5d1f0b4f972c8552faa475}{brisky-events} & 88.88 & 6 & 598  & 2 & 64 & 66 / 598 & 11.04 \\
18 & \href{https://github.com/c0bra/markdown-resume-js/tree/d43da4946c45fe3d6898672fe8b61b2d5ccaa8a9}{resume-js} & 97.73 & 14 & 648  & 3 & 48 & 51 / 648 & 7.87 \\
19 & \href{https://github.com/twilio/twilio-cli/tree/bcc0669d16d5c0debca4b7547866da71dac48fdd}{twilio-cli} & 97.49 & 16 & 308   & 5 & 17 & 22 / 308 & 7.14 \\
20 & \href{https://github.com/isellsoap/prettier-stylelint/tree/4e92b0fece371c054fe2df5f9e2c80e2b28032ff}{pret-temp} & 98.08 & 14  & 496 & 4 & 20 & 24 / 496 & 4.84 \\
21 & \href{https://github.com/alexjeffburke/relivestyle/tree/603acf4fc8178e30db677d0b9a8c653072dbeb96}{relivestyle} & 93.37 & 11 & 464 & 1 & 18 & 19 / 464 & 4.09 \\
22 & \href{https://github.com/APCOvernight/apc-build/tree/79273e6e40fe9eff9c22b2316b9476b754da7bd1}{apc-build} & 100 & 26  & 1,252  & 2 & 49 & 51 / 1,252 & 4.07 \\
23 & \href{https://github.com/albatrosary/generator-angular-eggs/tree/531395c4aa6945c75cf7f76c2a1567f19947de10}{gen-eggs} & 93.5 & 9  & 879  & 3 & 31 & 34 / 879 & 3.87 \\
24 & \href{https://github.com/codenautas/qa-control-server/tree/10fca9527bbf1705fb89b53f2541c62fd35b7f81}{qa-server} & 79.1 & 13  & 545  & 6 & 13 & 19 / 545 & 3.49 \\
25 & \href{https://github.com/hoodiehq/hoodie/tree/bd1354f10b2bcad154f24c3b35865ce3e05bd83a}{hoodie} & 100 & 24  & 941  & 4 & 29 & 33 / 941 & 3.5 \\
26 & \href{https://github.com/easy-team/res/tree/bf15ea88a78cae938fdaa4065cf8a055d69f33dc}{res} & 100 & 7 & 476  & 2 & 14 & 16 / 476 & 3.36 \\
27 & \href{https://github.com/hubcarl/easywebpack-vue/tree/b77f5f80615b353279da2d9183b6209b321bda18}{webpack-vue} & 100 & 5 & 1,434 & 3 & 44 & 47 / 1,434 & 3.28 \\
28 & \href{https://github.com/amokrushin/iamtest/tree/e63f0c75522d502f05f54d13d1690dc4da924b7d}{iamtest} & 90.1 & 14 & 600  & 3 & 16 & 19 / 600 & 3.17 \\
29 & \href{https://github.com/amokrushin/am-node-tape-runner/tree/e63f0c75522d502f05f54d13d1690dc4da924b7d}{am-runner} & 90.1 & 14 & 600  & 3 & 16 & 19 / 600 & 3.17 \\
30 & \href{https://github.com/Brightspace/images-to-variables/tree/cb2eb29640f116a891e1b17793f3e971e103d385}{img-variables} & 92.59 & 7 & 422 & 1 & 12 & 13 / 422 & 3.08 \\
31 & \href{https://github.com/vigour-io/brisky/tree/26113936272fc0e51f896dede83a006d3739fbb6}{brisky} & 92 & 10  & 443  & 1 & 10 & 11 / 443 & 2.48 \\
32 & \href{https://github.com/darlanmendonca/generator-bitch/tree/e4cce3142d733f787cc2db6b887d1ebfc654a350}{gen-bitch} & 99.52 & 4  & 505  & 1 & 11 & 12 / 505 & 2.38 \\
33 & \href{https://github.com/vigour-io/brisky-observable/tree/7b0019c4e3d30b145b45ae2dec42e5736cf28142}{observable} & 92.84 & 4  & 605  & 1 & 11 & 12 / 605 & 1.98 \\
34 & \href{https://github.com/vigour-io/state/tree/0060872464ee032127adea19e19f4b92587fe93c}{vigour-state} & 99.77 & 5 & 609  & 1 & 11 & 12 / 609 & 1.97 \\
35 & \href{https://github.com/vigour-io/brisky-class/tree/bb0e470d8aa5d81a4f12de65aa6dd254113846cd}{brisky-style} & 100 & 4  & 615  & 1 & 11 & 12 / 615 & 1.95 \\
36 & \href{https://github.com/vigour-io/brisky-style/tree/7b88d8f273a2efafe06a9c4bc3d53b7dc399dca6}{brisky-class} & 100 & 2  & 615  & 1 & 11 & 12 / 615 & 1.95 \\
37 & \href{https://github.com/fauxsoup/otp-js/tree/4f6729b0d187b7c0e013b4c39de62b373ef4e56c}{open-telecom} & 87.75 & 10  & 403  & 3 & 2 & 5 / 403 & 1.24 \\
38 & \href{https://github.com/spidercpsf/svgo/tree/c3be579af0a07624be6406a5dda5d6853ff83066}{svgo} & 84.6 & 10  & 411  & 3 & 2 & 5 / 411 & 1.22 \\
39 & \href{https://github.com/yadickson/generator-ajslib/tree/6a07944fac11b607424018d4d78cf5bcac799215}{hoodie-server} & 100 & 14 & 479 & 2 & 0 & 2 / 479 & 0.42 \\
40 & \href{https://github.com/yadickson/generator-ajslib/tree/6a07944fac11b607424018d4d78cf5bcac799215}{gen-ajslib} & 100 & 10  & 474  & 1 & 1 & 2 / 474 & 0.42 \\
41 & \href{https://github.com/TestArmada/magellan-seleniumgrid-executor/tree/5c3d11ebfe575d9251fe41997435179c90f76ea8}{tst-executor} & 98.05 & 7  & 365 & 1 & 0 & 1 / 365 & 0.27 \\
42 & \href{https://github.com/taoyuan/mqttoxy/tree/3b08f6070bf8b840a66ff01c2d340af69468c11e}{mqtter} & 90.32 & 7 & 420 & 1 & 0 & 1 / 420 & 0.24 \\
43 & \href{https://github.com/ladjs/lad/tree/bc0f9722c3dd0af690be8a34e02751ce091a7891}{lad} & 76 & 14  & 568  & 1 & 0 & 1 / 568 & 0.18 \\
44 & \href{https://github.com/jfmbrennan/generator-ngbp-module/tree/534d854532c89aa706680d0c2f52464a5653bdbd}{gen-module} & 100 & 11  & 1,217 & 1 & 1 & 2 / 1,217 & 0.16 \\
45 & \href{https://github.com/weblogixx/generator-react-webpack-alt/tree/8b233731dc899d88945372bc05d601715bb822d5}{gen-alt} & 96.06 & 4  & 1,138 & 1 & 0 & 1 / 1,138 & 0.09 \\
46 & \href{https://github.com/hubcarl/easywebpack-js/tree/f45f5f03f486e532d1f7b1e480dc374cb567e37a}{easywebpack-js} & 100 & 2 & 1,523  & 1 & 0 & 1 / 1,523 & 0.07 \\
\midrule
 & Total & - & 462  & 28,084 & \totalNbBloatedDirectDeps & 4,167 & 4,287 / 28,084 & - \\
% \botrule
\end{tabular*}
% \footnotetext[1]{Number of indirect bloated dependencies that are caused by direct bloated dependencies}
\end{table}

% one direct leads to significant bloats
Out of \nbPackagesDirectBloated packages, 40 (87.0\%) experience the removal of indirect dependencies as a direct consequence of removing bloated direct dependencies.
We see a significant cascading effect occurs when indirect dependencies are entirely associated with bloated direct dependencies. 
For example, in the package \href{https://github.com/Gleider/podcast-search/tree/8a79ed1ed37c9d345a832e7e2dfef823e0e200c7}{\texttt{podcast-search}} at row 2,
using the direct-only debloating strategy to remove the bloated direct dependency \texttt{npm} leads to the removal of 679 indirect dependencies.
In this case, all the indirect dependencies are exclusively associated with the bloated direct dependency, while the non-bloated direct dependency does not pull any indirect dependencies into the package.
This straightforward one-line deletion maintenance activity results in a notable overall removal of 680 out of 681 (99.8\%) dependencies, demonstrating the significant impact of addressing direct bloated dependencies.

% direct leads to > 10 indirects
Such cascading effects are not isolated.
Focusing on the column \texttt{\#BD$\rightarrow$I}, we observe that out of the \nbPackagesDirectBloated packages, 36 (76.6\%) packages immediately delete more than 10 indirect bloated dependencies debloating the direct dependencies, indicating that 
For example, in the package \href{https://github.com/moneybutton/bsv-mnemonic/tree/6c73e2a47aba7aad6758aa820c9b59faa47ef0fc}{\texttt{bsv-mnemonic}} at row 6, the removal of 2 direct bloated dependencies results in the additional removal of 309 indirect dependencies.
The removal of these indirect dependencies accounts for a reduction of more than half (51.3\%) of the dependency tree of \texttt{bsv-mnemonic}, when simply deleting 2 direct dependencies declarations from the \package file.
This observation highlights the cascading effect of bloated dependencies within software dependency trees, and the amplified effects of taking care of bloat directly in the \package file.

% 6 cases without impact on indirect, the need for thoroughly debloat indirect
Although this question focuses on removing direct bloated dependencies, we observe that indirect bloated dependencies can also exist independently of direct bloated dependencies, often introduced through non-bloated direct dependencies.
\edit{These dependencies account for 83.1\% (21,159 out of 25,446) of all indirect bloated dependencies, indicating that the majority of indirect bloat are not introduced through direct bloated dependencies, but instead originate from non-bloated direct dependencies.
}
For example, the package \href{https://github.com/aliaksandr-masteit challenging to r/generator-node-lib/tree/de5890cf0e3b60d389008ae045993364c59cae25}{\texttt{generator-node-lib}} contains 798 indirect bloated dependencies despite having no direct bloated dependencies.
Similarly, \href{https://github.com/easy-team/easywebpack-js/tree/f45f5f03f486e532d1f7b1e480dc374cb567e37a}{\texttt{easywebpack-js}} (at row 46 in \autoref{table:direct_bloated}) includes 1,521 dependencies in its tree, with only one direct bloated dependency that does not contribute to the removal of any indirect dependencies. 
Among its indirect dependencies, 80.7\% (1,227/1,521) are bloated and stem from its single non-bloated direct dependency. 
These findings demonstrate that indirect bloated dependencies can propagate through non-bloated direct dependencies, independent of direct bloat, highlighting the challenge of slimming dependency trees.

Our observation is evidence that bloated dependencies can appear anywhere in the dependency tree, representing a major challenge for developers, as there is currently no support for identifying and removing them effectively \citep{kabir2022developers}.
By focusing on direct bloated dependencies through the direct-only debloating strategy, we observe notable reductions in dependency tree size. 
However, fully mitigating dependency bloat requires additional efforts to target indirect dependencies, which currently lack native tool support within the \npm ecosystem.

\begin{answerbox}
\textbf{Answer to RQ2:}
Trace-based analysis using \depPrune identifies \nbPackagesDirectBloated packages in our dataset that have at least one direct bloated dependency. 
Removing the declaration of the \totalNbBloatedDirectDeps direct bloated dependencies from the \package files of these packages immediately removes \totalNbBloatedInrectDepsByDirect indirect dependencies as well.
Maintaining the \package to ensure that it does not contain direct bloated dependencies is a healthy engineering practice, which needs to be better supported by package managers.
\end{answerbox}

\subsection{RQ3: \RQthree}
To evaluate the effectiveness of our trace-based analysis in detecting bloated dependencies, we compare it against two state-of-the-art techniques: a dynamic coverage-based approach and a static analysis approach, across our dataset of \nbPackages \commonjs packages.
By analyzing overlaps and gaps in the detection of bloated dependencies, we highlight the strengths and potential areas for improvement in bloat detection techniques, aiming to provide valuable insights for enhancing the debloating of \commonjs packages.

\begin{figure}
    \centering
    \captionsetup{font=scriptsize}
    % Subfigure (a)
    \begin{subfigure}{0.5\textwidth}
        \centering
        \captionsetup{font=scriptsize}
        \includegraphics[width=\textwidth]
        {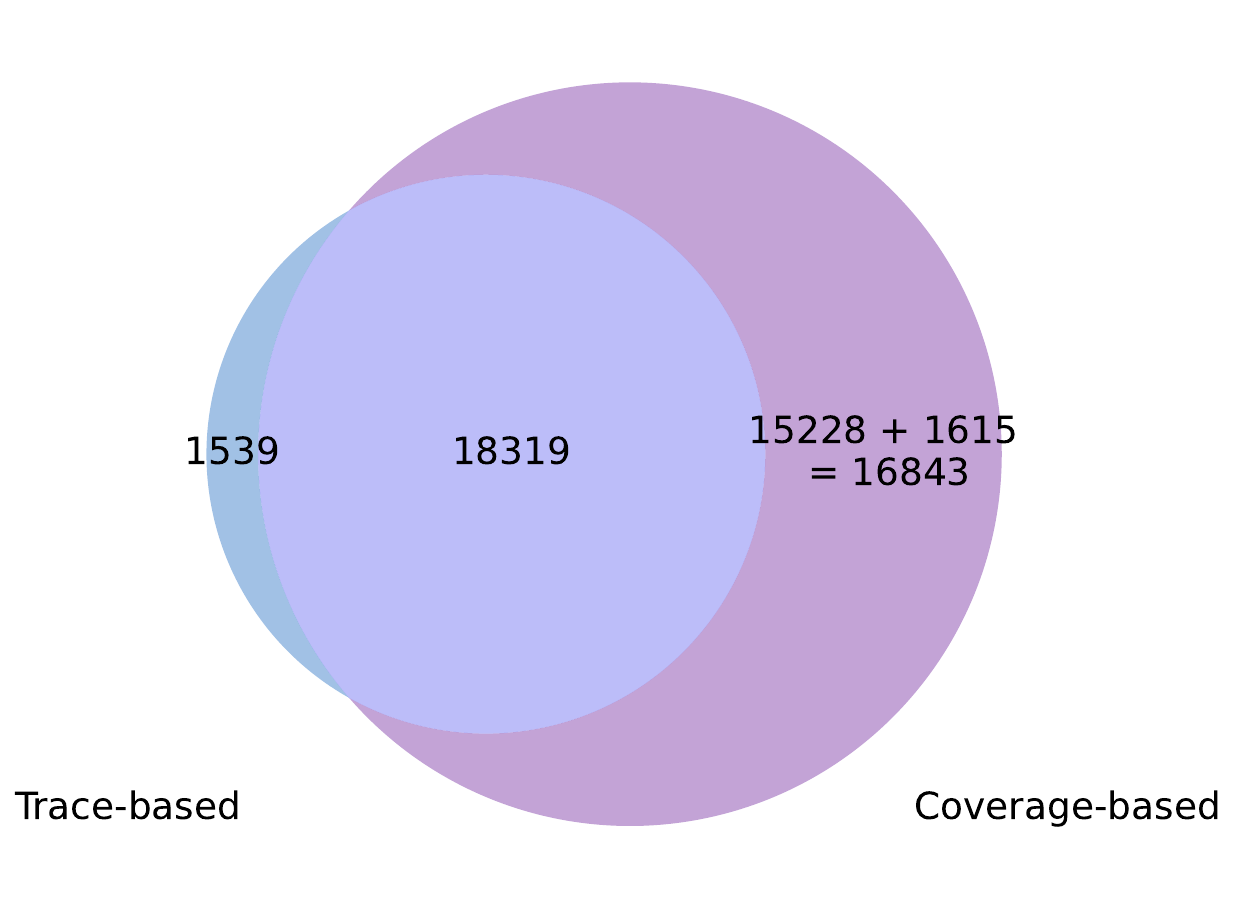} % Replace with your image
        \caption{Total bloated dependencies identified in 77 packages, compared between trace-based analysis using \depPrune and coverage-based analysis using \stubbifier.}
        \label{fig:RQ3_Venn-a}
    \end{subfigure}
    \hfill
    % Subfigure (b)
    \begin{subfigure}{0.45\textwidth}
        \centering
        \captionsetup{font=scriptsize}
        \includegraphics[width=\textwidth]{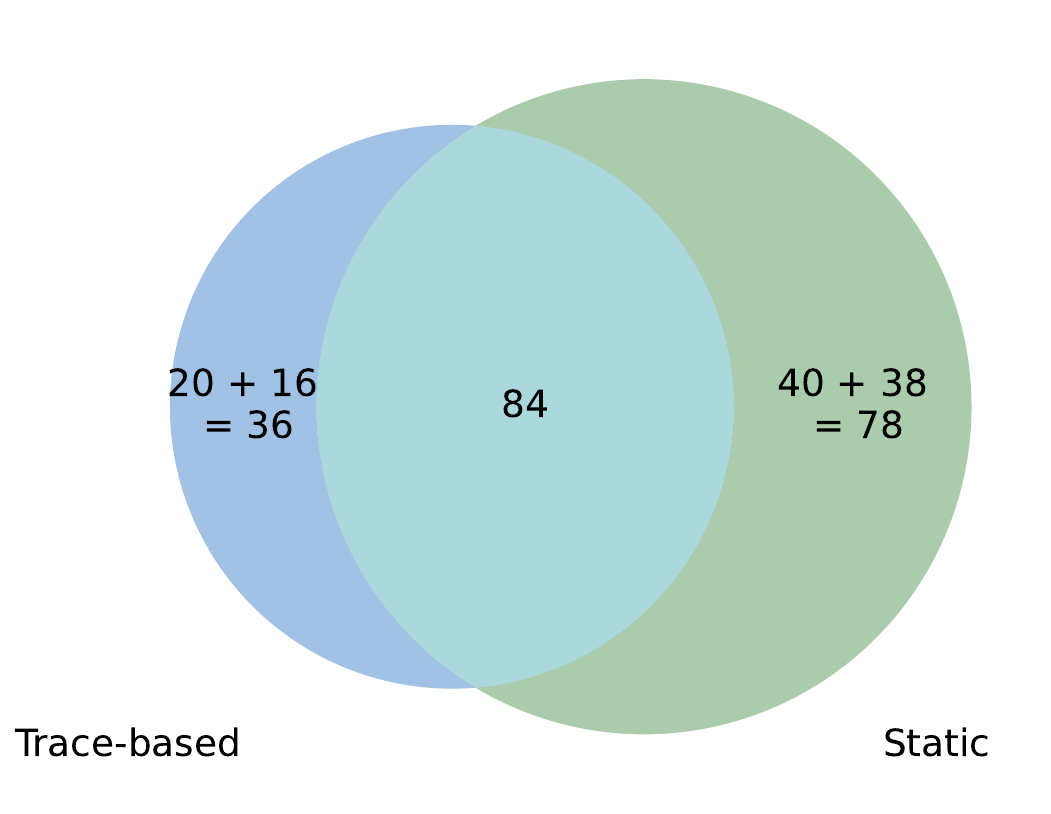} % Replace with your image
        \caption{Direct bloated dependencies identified in 91 packages, compared between trace-based analysis using \depPrune and static analysis using \depCheck.}
        \label{fig:RQ3_Venn-b}
    \end{subfigure}
    \vspace{1em}
    \caption{Within total bloated dependencies identified in \nbPackagesStubbifier packages between trace-based analysis using \depPrune and coverage-based analysis using the adapted \stubbifier,
    18,319 dependencies are identified as bloated by both analyses.
    Trace-based analysis identifies 19,858 (18,319 + 1,539) bloated dependencies, while coverage-based analysis identifies 35,162 (18,319 + 16,843) bloated dependencies.
    Note that no incorrectly identified bloated dependencies are identified by trace-based analysis, whereas coverage-based analysis misclassifies 15,228 dependencies as bloated and misses 1,539 dependencies that are identified by trace-based analysis. 
    Additionally, 1,615 dependencies are classified as dynamically resolvable.
    Within direct bloated dependencies identified in \nbPackages packages compared between trace-based analysis using \depPrune and static analysis using \depCheck,
    84 dependencies are identified as bloated by both analyses.
    Trace-based analysis identifies 120 (84 + 36) bloated dependencies, while static analysis identifies 162 (84 + 78) bloated dependencies.
    Note that trace-based analysis potentially misclassifies 20 dependencies as bloated and misses 38 dependencies, while static analysis misclassifies 40 dependencies as bloated dependencies and overlooks 16 dependencies.
    }
    \label{fig:RQ3_Venn}
\end{figure}

% overall
\subsubsection{Comparison with Dynamic Analysis}
In the first section of our comparative study, we compare our trace-based approach using \depPrune, to the state-of-the-art dynamic coverage-based bloat analysis using the adapted \stubbifier,  for the detection of all (both direct and indirect) bloated dependencies.
As outlined in \Cref{subsec:ptcl3}, we run both \depPrune and the adapted \stubbifier on our dataset of \nbPackages \commonjs packages.
Out of the \nbPackages packages, the adapted \stubbifier successfully runs on \nbPackagesStubbifier.
For the remaining 14 packages, it encounters compatibility errors or terminates during execution due to a failure in parsing a specific file.
Consequently, 14 packages are excluded, and we compare the precision of both approaches on the remaining \nbPackagesStubbifier packages.

% overall coverage-based
Across the \nbPackagesStubbifier packages, the coverage-based analysis using the adapted \stubbifier identifies a total of 35,162 unaccessed dependencies.
To verify whether these dependencies are truly bloated, we remove them, rebuild the packages, and run their tests.
Among the dependencies identified as unaccessed, 19,934 (56.7\%) can be safely removed while preserving the original package behavior ($18,319 + 1,615$ in \autoref{fig:RQ3_Venn-a}).
However, 15,228 (43.3\%) of these dependencies result in test failures when removed, indicating that a significant rate of them are misclassified.

% overall trace-based
In contrast, using trace-based analysis with \depPrune, we identify a total of 19,858 unaccessed dependencies, and all of them can be safely removed ($18,319 + 1,539$ in \autoref{fig:RQ3_Venn-a}).
Notably, 1,539 dependencies identified as bloated by the trace-based analysis are not detected by coverage-based analysis.
This result demonstrates the capability of the trace-based approach to capture dependencies overlooked by the coverage-based method.

We observe that 1,615 dependencies are identified as bloated by the coverage-based analysis using \stubbifier, but classified as non-bloated by our trace-based approach using \depPrune. 
Interestingly, these dependencies can be removed without affecting functionality.
This behavior arises because, while accessed by the OS file system during execution, these indirect dependencies (classified as dynamically resolvable dependencies in \autoref{fig:RQ3_Venn}) are replaced by alternative versions due to \javascript's module resolution mechanism.
Specifically, after their removal, the \javascript engine resolves and uses an available version of the same dependency from an ancestor directory in the file system hierarchy.

For example, in the package \href{https://github.com/ciena-blueplanet/eslint-plugin-ocd/tree/d22871dc398a9addd173a1963675c1757ea27384}{\texttt{eslint-plugin-ocd}}, two versions of the dependency \texttt{resolve-from} are resolved. 
Version 2.0.0 is installed in the root \nodemodules folder, while Version 1.0.1 resides in the \nodemodules folder of its parent dependency \texttt{require-uncached}, which specifies a version incompatible with Version 2.0.0. 
Using \depPrune, Version 1.0.1 is correctly identified as non-bloated (rows 1, 2, and 3 in \autoref{lst:rq3_example_before}), whereas \stubbifier classifies it as bloated.
After removing Version 1.0.1 of \texttt{resolve-from}, the package \texttt{eslint-plugin-ocd} still passes all its tests. 
Upon tracing its execution, we observe that the OS reports an error when attempting to access Version 1.0.1 (row 1 in \autoref{lst:rq3_example_after}); however, it successfully resolves to Version 2.0.0 in the root \nodemodules folder (row 2 in \autoref{lst:rq3_example_after}).
This dynamic resolution enables the package’s functionality to remain intact despite the removal.

In this scenario, the removed dependency is replaced by another version that continues to fulfill the package's functional requirements during testing. 
We believe this behavior does not present misclassification of \depPrune but rather reflects a specific scenario enabled by the \javascript engine's dynamic resolution capabilities. 
While this mechanism avoids immediate functional issues, removing such dependencies can still introduce risks such as compatibility problems or breaking changes if future updates lead to inconsistent behavior across versions.

\begin{lstlisting}[caption={Example of an indirect dependency classified as non-bloated by \depPrune. The dependency \texttt{resolve-from} is accessed via the \texttt{node\_modules} folder of its parent dependency \texttt{require-uncached}, as captured during the trace-based analysis.},
  label={lst:rq3_example_before},
  breaklines=true,
  basicstyle=\scriptsize,
  breakatwhitespace=true]
1. 816 openat(AT_FDCWD, '/disk/eslint-plugin-ocd/%*\fboxsep 1pt \colorbox{yellow!20}{node\_modules/require-uncached/node\_modules/resolve-from}*)
   /package.json', O_RDONLY|O_CLOEXEC) = 22
2. 816 openat(AT_FDCWD, '/disk/eslint-plugin-ocd/%*\fboxsep 1pt \colorbox{yellow!20}{node\_modules/require-uncached/node\_modules/resolve-from}*)
   /index.js', O_RDONLY|O_CLOEXEC) = 22
...
\end{lstlisting}

\begin{lstlisting}[caption={Example of an indirect dependency classified as non-bloated by \depPrune but dynamically resolved to an alternative version after its removal. The dependency \texttt{resolve-from}, initially accessed via the \texttt{node\_modules} folder of \texttt{require-uncached}, is no longer found after its removal (ENOENT). Instead, the \javascript engine resolves to another version of \texttt{resolve-from} in the root \texttt{node\_modules} folder, maintaining the package's functionality.},label={lst:rq3_example_after},breaklines=true, basicstyle=\scriptsize, breakatwhitespace=true]
1. 1090 openat(AT_FDCWD, '/disk/eslint-plugin-ocd/%*\fboxsep 1pt \colorbox{yellow!20}{node\_modules/require-uncached/node\_modules/resolve-from}*)
   /package.json', O_RDONLY|O_CLOEXEC) = -1 %*\fboxsep 1pt \colorbox{yellow!20}{ENOENT (No such file or directory)}*)
2. 1090 openat(AT_FDCWD, '/disk/eslint-plugin-ocd/ %*\fboxsep 1pt \colorbox{yellow!20}{node\_modules/resolve-from}*)
   /package.json', O_RDONLY|O_CLOEXEC) = 22
3. 1090 openat(AT_FDCWD, '/disk/eslint-plugin-ocd/%*\fboxsep 1pt \colorbox{yellow!20}{node\_modules/resolve-from}*)
 /index.js", O_RDONLY|O_CLOEXEC) = 22
...
\end{lstlisting}

% missed dependencies in coverage-based
On the other hand, the coverage-based analysis misses 1,539 bloated dependencies by failing to detect them in certain scenarios.
This occurs due to the filtering mechanisms in \stubbifier, which ignore specific files, such as processing only .js files and deliberately ignoring all \texttt{@babel} dependencies\footnote{\href{https://github.com/emarteca/stubbifier/blob/master/stubbifyRunner.ts}{Coverage-based analysis filtering files.}}. 
In contrast, our trace-based analysis tracks all runtime dependencies without ignoring any specific file type and can capture these missed dependencies.

\bigskip
We see that a significant proportion ($15,228$ or 43.3\%) of the dependencies identified by the coverage-based analysis are misclassified as bloated.
These dependencies include cases where modules are loaded but are determined to be non-executed, causing them to be marked as unaccessed. 
To investigate the underlying reasons for these misclassifications generated by coverage-based analysis, we conduct a manual analysis of a random selection of 100 dependencies, using a 95\% confidence level for our sampling.
% 9.77\%
\edit{The margin of error for this sample is 9.77\%.}

For each selected dependency, we remove it from the package, rerun the tests, and analyze the test reports to identify errors caused by the missing dependency.
This qualitative analysis identifies three reasons why certain runtime dependencies, which are marked as unaccessed in the coverage-based analysis, cannot actually be removed. 
We discuss these reasons, along with the corresponding number of cases identified for each category.

\emph{Neglected code outside of functions (77/100)}: 
As discussed in \Cref{subsec:stateofart}, the state-of-the-art coverage-based analysis tool deems a module unused if none of its functions' instructions are covered.
However, in \javascript, executable code can exist outside functions \citep{maras2016secrets}, or a module can export only data rather than functions.
Hence, assessing function execution alone is insufficient when evaluating file access through coverage-based analysis; all executable code within a file must be considered.
Neglecting this aspect can result in the misclassification of a file's usage. 
In contrast, trace-based analysis mitigates this issue by monitoring file system interactions directly, ensuring that all executable code is captured during runtime, regardless of its location within the file.

\emph{Mislabeled executed functions (20/100)}:
Coverage-based analysis struggles to accurately detect function executions due to its reliance on static instrumentation and straightforward coverage reports, which are limited in capturing dynamic and non-linear execution flows. 
Constructs like Immediately-invoked function expressions (IIFEs), nested functions, inline functions, and asynchronous behaviors of functions introduce complexity, as their execution often depends on runtime conditions that static instrumentation cannot fully trace.
For example, functions executed in a delayed manner, such as callbacks or functions returned and later invoked, can appear unexecuted in coverage reports.
This is because coverage data is typically generated immediately after test execution, potentially missing code executed asynchronously beyond that point.
This limitation highlights the challenges of analyzing modern JavaScript patterns, which frequently rely on dynamic function invocations and non-blocking execution mechanisms.
Our results strongly support previous studies \citep{richards2010analysis, esben2017} and provide further evidence that the dynamic behaviors of \javascript pose significant challenges for testing and debugging in \nodejs \citep{wang2017comprehensive, sun2019reasoning, hashemi2022empirical}.
In contrast to coverage-based analysis, trace-based analysis has the capability to capture every executed file, regardless of the timing or context of their execution, making it a more reliable method for identifying runtime dependencies. 
Our finding underscores the need for comprehensive testing reports in \commonjs packages to enhance the accuracy of coverage-based analysis.

\emph{Required modules of data files (3/100):} 
We find that some modules only import data from a dependency, which is typically stored in a JSON file.
However, such JSON files are not reported in a test coverage report.
Consequently, coverage-based analysis marks the file and the corresponding dependency as unaccessed, although it is necessary to run the package.
In contrast, trace-based analysis can recognize these file accesses and report such dependencies as non-bloated.
This finding suggests that test coverage report tools should take data files into consideration for a more comprehensive report.

\subsubsection{Comparison with Static Analysis}
As the second section of our comparative analysis, we compare trace-based analysis using \depPrune, with static analysis using \depCheck, a standard \npm tool for finding unused direct dependencies \citet{depcheckgitaddress}.
After running both tools on our dataset of \nbPackages packages, we confirm that both \depPrune and \depCheck successfully execute on all packages.

Across the \nbPackages packages, \depCheck identifies \directBloatedByDepcheck direct bloated dependencies.
To verify these detections, we remove each identified dependency from the package, rebuild the package, and rerun the tests.
Our results, presented in \autoref{fig:RQ3_Venn-b}, reveal that both \depCheck and \depPrune successfully identify 84 dependencies that can be safely removed without affecting functionality.

Additionally, \depCheck identifies 78 (40 + 38) dependencies as bloated.
However, these dependencies break the tests after each removal.
By analyzing each breaking error, we find that 40 dependencies are dynamically generated at runtime when imported.
For example, in the package \href{https://github.com/screwdriver-cd/screwdriver/tree/cc915df162b3aa8117ad8f2294ced64c7f75513d}{\texttt{screwdriver}}, dependencies are imported dynamically during runtime on a large scale, where 10 of them are wrongly detected as bloated by \depCheck.
This result underscores the error-prone nature of static code analysis approaches when applied to a scripting language such as \javascript, a sentiment supported by prior research \citep{kabir2022developers}.
In addition, 38 dependencies identified as bloated by \depCheck are testing-related development dependencies misused by developers as runtime dependencies in the \package file. 
Differ from the dependencies discussed in \autoref{table:bloated_reason}, these dependencies are captured as used and identified as non-bloated by \depPrune, indicating their usage during testing validation.
These findings further emphasize the prevalence of misused development dependencies in \commonjs packages.

On the other hand, 36 direct dependencies identified as bloated by \depPrune are detected as non-bloated by \depCheck.
Among these, 20 dependencies are related to command-line tasks, which are not exercised during testing and may present misclassification under \depPrune's methodology.
The remaining 16 missed dependencies are cases where \depCheck failed to detect non-usage. 
For example, in the package \href{https://github.com/jclo/umdlib/blob/572af0377f2194339eab79554c6b7e0bb7160ffe/bin/umdlib.js}{\texttt{umdlib}}, 2 dependencies (\texttt{nopt} and \texttt{path}) are required in the same manner within a single file.
However, \depCheck identifies \texttt{path} as bloated while overlooking \texttt{nopt}.
This result shows that static analyses struggle with unconventional or complex dependency importing patterns in \javascript, even if the dependencies are not dynamically generated during runtime.

\bigskip
\begin{answerbox}
\textbf{Answer to RQ3:}
The state-of-the-art dynamic coverage-based analysis tends to misclassify dependencies as bloated due to the heavy reliance on the coverage report tool.
On the other hand, static analysis fails to detect dynamically generated dependencies during runtime and is error-prone when analyzing complex dependency usage patterns in \commonjs 
 packages.
In contrast, our trace-based approach effectively addresses these challenges by leveraging runtime execution monitoring to capture actual dependency usage.
\end{answerbox}

\subsection{Performance}
\depPrune successfully executes and analyzes all \nbPackages packages in the dataset, using their test suites as workloads.
All identified bloated dependencies can be safely removed while preserving the packages' functionality during evaluation.
The additional time required by \depPrune, measured as a multiple of the original workload execution time, varies from 0.2x to 27.4x, with a median of 1.4x.
For example, in the package \texttt{fastify-axios}, the original workload takes 3.5s while running \depPrune on this package takes 4.2s, resulting in the additional 0.7s (0.2x) of the original workload.
In addition, 22 packages have this overhead lower than 1x, demonstrating minimal additional time required for our analysis.
\edit{While 3 packages have an overhead higher than 10x, this occurs primarily in packages with a small number of test cases, where the original test execution time is very short, making the additional analysis time appear relatively large.}
Notably, all packages with more than 100 test cases have an overhead lower than 2x.
For example, the package \href{https://github.com/cedric05/cli/tree/54079eadb5ba8f4c7e45c2876ea589e5e4844342}{cedric05-cli}, which has 4,396 test cases, has an overhead of only 1.04x.
These results highlight that the overhead introduced by \depPrune is generally low, making it a practical and efficient solution for dependency analysis with manageable runtime overhead.

\edit{\subsection{\depPrune in the Development Lifecycle}
\depPrune debloats both direct and indirect dependencies. To do so, our tool automatically rewrites \package and \packagelock, enabling developers to build and install a package with no bloated dependencies.
While \packagelock is typically considered transient and may be regenerated during development, it plays a critical role in shaping the installed dependency tree.
To prevent bloated indirect dependencies from reappearing, we recommend running \depPrune before each release.
This ensures that the deployed environment includes only the necessary dependencies, resulting in a leaner dependency tree, without requiring changes from the \npm package manager.
This is particularly relevant in workflows that use bundlers (e.g., Webpack\footnote{\href{https://webpack.js.org/concepts/}{\texttt{Webpack: a bundler for modern \javascript applications}}}), where the dependency tree directly impacts bundle size.
}

%%=============================================================%%

%%=============================================================%%
\section{Threats to Validity}
Our results reveal that trace-based analysis, which monitors the interactions between files and the OS file system, is reliable for debloating \npm dependencies.
In this section, we discuss the threats to the validity of our results.

\subsection{Internal Validity}
A key threat to the internal validity of our study is our reliance on dynamic analysis to identify bloated dependencies.
While our approach demonstrates that dynamic analysis is an effective method for identifying bloated dependencies in \commonjs packages, the accuracy of our findings relies on the comprehensiveness of the workload used.
To mitigate this threat in our evaluation, we curate a dataset of packages with at least 70\% test coverage, achieving a median test coverage of \testCov. 
This threshold helps ensure that a significant portion of the package's functionality is exercised, thus providing a reliable workload for our dynamic analysis. 
In a practical setting, packages with incomplete or missing test suites can still be analyzed using any other workload that exercises the package's functionality.
This includes integration tests, functional validation, or real-world scenarios where the package is executed in a typical environment.
Additionally, designing tailored workloads for specific applications can further address this limitation.
Future work could focus on enhancing our approach by complementing dynamic analysis with additional or alternative workloads to achieve a more complete understanding of dependency usage.

\subsection{External Validity}
% diverse packages
The external validity of our study concerns the extent to which our findings can be generalized beyond the specific context of our experiments. 
The diverse nature of the packages in our dataset helps mitigate threats to external validity.
Our dataset includes \nbPackages real-world, open-source, well-tested packages from the \nodejs and \commonjs ecosystems.
These packages vary significantly in size, the complexity of their dependency trees, and the application domains they cover.
They represent server-side operations such as building, linting, and live-reloading, as well as development frameworks, SVG file manipulation, and podcast registry data retrieval.
This diversity strengthens our claim that the results reflect broader trends within the \nodejs and \commonjs ecosystems.
In addition, the core principle of our trace-based analysis — monitoring system calls and file interactions at the operating system level — ensures independence from higher-level tools and frameworks.
This makes the approach inherently scalable across different operating systems, allowing it to be applied in other environments.

\subsection{Construct Validity} 
% misclassfication
A potential threat to construct validity arises from the misuse of dependencies in the \package file, particularly testing-related dependencies that are declared as runtime dependencies.
While our approach identifies dependencies accessed during workload execution as non-bloated, misused testing dependencies are retained as they are treated as required runtime dependencies.
% developers should carefully classify them
Our finding emphasizes the importance of accurate dependency declarations for improving debloating results.
To accurately reflect the set of runtime dependencies in a production environment, developers need to carefully classify development and runtime dependencies in the \package file.

%%=============================================================%%

%%=============================================================%%
\section{Related Work}

\subsection{\javascript Debloating}
Code bloat is an issue in both client-side \citep{vazquez2019slimming, kupoluyi2021muzeel, malavolta2023javascript} and server-side \citep{koishybayev2020mininode, turcotte2021stubbifier} \javascript applications.
Existing work on the client side has focused on code size minimization, with much has been accomplished through both static and dynamic analysis.
Static analysis can address client-side bloat through the capabilities of tree-shaking in web bundlers \citep{treeshaking, pavic2021methods}.
The effectiveness of tree-shaking in eliminating code bloat stems from the static import mechanism on the client side, i.e. ES6 \citep{MDNimport}, allowing for the determination of imported dependency code before program execution.
Regarding dynamic analysis, \citet{vazquez2019slimming} employ a method to instrument functions within \texttt{.js} files and analyze runtime executions through tests to suggest unused functions for removal.
\citet{kupoluyi2021muzeel} propose \texttt{MUZEEL}, a technique to identify unused functions by triggering associated events to emulate user interactions, considering functions not called within any event as eliminable.
Another work from \citet{malavolta2023javascript} delves deeper into the runtime overhead of the unused functions.
They demonstrate the positive impact of removing unused functions in terms of energy consumption, performance, network usage, and resource utilization.

Debloating in server-side \javascript primarily addresses security concerns with the aim of reducing the attack surface \citep{koishybayev2020mininode, turcotte2021stubbifier, ye2021jslim}.
\citet{koishybayev2020mininode} contribute to reducing the attack surface of \npm packages and their dependencies by eliminating code at both the function and file levels.
Their approach involves constructing function-level and file-level call graphs to identify unused functions or files, accompanied by a comprehensive static analysis that addresses challenges related to complex module imports and exports.  
\citet{ye2021jslim} collect information about vulnerabilities and build a function-level call graph to establish the relationship between functions and vulnerabilities.
Both \citet{koishybayev2020mininode} and \citet{ye2021jslim} acknowledge the challenge of identifying dynamically loaded modules through static analysis, which is natural for a scripting language, as we illustrated.
In contrast, \citet{turcotte2021stubbifier} perform a dynamic analysis on 15 \npm packages to detect potentially unused functions and files.
In their approach, they replace them with context-called stubs that can dynamically fetch and execute the original code if needed.

Our work differs from the above studies in several aspects.
We conduct dynamic analysis on \commonjs packages, which include dynamically loaded modules mentioned in previous research \citep{koishybayev2020mininode, ye2021jslim}.
We evaluate our approach on a much larger dataset, using packages with high test coverage (median \testCov).
Our focus is on removing entire dependencies from the dependency tree of a package, leaving the original package source code unchanged, a novel aspect not previously explored in \javascript.
Our approach can aid developers in reducing the maintenance costs associated with bloated dependencies, such as notification fatigue.

\subsection{Applications Debloating}
Comprehensive code usage analysis forms a critical foundation for software debloating, presenting challenges for both static and dynamic analysis in applications.
Static analysis faces difficulties in accurately predicting code usage due to the dynamic nature of user inputs, while dynamic analysis encounters issues with excessive code execution overhead. 
% One sentence to summarize
Given these challenges, research in application debloating has proposed specific analysis methods to identify code usage in terms of the programming language's features and then debloat code at different levels of granularity.
Meanwhile, evaluating impacts on code size and on security are often key measurements of debloating effectiveness.

% Less is more: Quantifying the security benefits of debloating web applications
\citet{azad2019less} present the first analysis of the security benefits of debloating web applications, focusing on PHP applications.
Their evaluation of function-level and file-level debloating strategies demonstrated notable reductions in size and cyclomatic complexity, as well as the removal of code associated with historical vulnerabilities and unnecessary external packages.
% Role Models: Role-based Debloating for Web Applications: dynamic, features
\citet{amin2023role} focus on clustering user roles by collecting similar usage behavior and removing unneeded features, then providing users with a custom debloated application that is tailored to their needs.
% AnimateDead: Debloating Web Applications Using Concolic Execution
\citet{azad2023animatedead} propose a hybrid approach for analyzing PHP module reachability based on concolic execution, aiming to bridge the gap between the scalability of dynamic debloating and the accuracy of static analysis.
% HODOR: Shrinking Attack Surface on Node.js via System Call Limitation
% DeView: Confining Progressive Web Applications by Debloating Web APIs：dynamic, API
\citet{oh2022deview} reduce PWA (progressive web application) attack surface by blocking unnecessary web APIs, employing record-and-replay web API profiling to identify needed web APIs and eliminating unnecessary entries through compiler-assisted debloating in the browser. 
\citet{wang2023hodor} developed the \texttt{HODOR} tool to minimize the attack surface of \nodejs applications with the goal of strengthening security against arbitrary code execution vulnerabilities.
They construct call graphs for \nodejs applications through a combination of static and dynamic call graph analyses.
Instead of eliminating \javascript source code, they focus on restricting system calls that are invoked by the code.
% Jshrink: In-depth investigation into debloating modern java applications
\citet{bruce2020jshrink} present JShrink, a bytecode debloating framework for modern Java applications, emphasizing dynamic language features and ensuring behavior preservation through regression testing.
% Efficiently Trimming the Fat: Streamlining Software Dependencies with Java Reflection and Dependency Analysis.
\citet{song2024efficiently} propose \texttt{Slimming}, which statically analyzes Java reflections commonly used in frameworks, aims to remove unnecessary dependencies and improve dependency management.
% Cesar's comprehensive, longitude, specification
\citet{soto2021comprehensive, soto2021longitudinal, soto2023automatic} extensively research on debloating dependencies under the Maven system.
They develop a tool to identify completely useless dependencies through static analysis, focusing on identifying unused dependencies through static analysis and exploring their prevalence, evolution, and maintenance costs.
They also propose techniques to optimize Java dependency trees by specializing dependencies based on actual usage.

Our work differs conceptually in both scope and methodology, addressing challenges specific to the dynamic and scripting nature of \javascript.
While static analysis is effective in statically typed languages like Java, \javascript's runtime behavior and the \commonjs module system complicate such approaches.
This dynamic nature, combined with features like \require's parameterized imports and runtime dependency generation, limits the applicability of static debloating techniques.
To address these challenges, we adopt a dynamic, trace-based approach that monitors OS file system interactions during runtime, effectively capturing dependencies missed by static analysis.
Our approach bridges a critical gap in dependency debloating research, offering specific solutions for \commonjs projects and providing insights into the unique dependency management challenges posed by the \npm ecosystem.
We believe these contributions distinguish our work from prior research and address a critical gap in dependency debloating for JavaScript projects. 

\subsection{\javascript Dependency Management}
Related studies on \javascript dependency management primarily focus on uncovering issues and assisting developers in improving dependency management practices, particularly within the \npm ecosystem.
% Dependency smells in Javascript projects.
\citet{jafari2021dependency} delve into identifying and quantifying dependency smells in \javascript projects, shedding light on security threats, bugs, dependency breakage, and maintenance challenges stemming from improper dependency management.
% An Empirical Analysis of Technical Lag in npm Package Dependencies
\citet{zerouali2019formal} introduce a technical lag metric for \npm package dependencies, revealing significant delays in updating due to dependency constraints, thereby impacting software development practices.
% Not All Dependencies are Equal: An Empirical Study on Production Dependencies in NPM
\citet{latendresse2022not} study the frequency of dependencies being released to production, revealing the challenges associated with both runtime and development dependencies.
% An empirical study of dependency downgrades in the npm ecosystem
\citet{cogo2019empirical} investigate dependency downgrades in \npm, focusing on the underlying reasons for these downgrades.
% Toward using package centrality trend to identify packages in decline
% Where to Go Now? Finding Alternatives for Declining Packages in the npm Ecosystem
\citet{mujahid2021toward, mujahid2023go} leverage package centrality trends to identify declining packages in \npm and suggest alternative packages based on dependency migration patterns and centrality trends.
% Dependency Update Strategies and Package Characteristics
% Characterizing Dependency Update Practice Of NPM, PyPI and Cargo Packages
In addition, the maintenance overhead associated with updating dependencies is a growing concern in dependency management.
\citet{javan2023dependency} and \citet{rahman2024characterizing} explore dependency upgrade strategies with the goal of helping developers avoid the risk of potentially high maintenance overhead when selecting dependencies.

These studies collectively highlight the challenges associated with importing dependencies, underscoring the significant maintenance costs incurred.
It is precisely these challenges that we address in our work.
Our research aims to assist developers in identifying unnecessary dependencies in their packages to effectively mitigate maintenance overheads.

%%=============================================================%%

%%=============================================================%%
\section{Conclusion}
% context 
In this paper, we conduct a comprehensive study on bloated dependencies in \nodejs packages that use the \commonjs module system.
% approach
We propose a trace-based approach, implemented in a tool \depPrune, which monitors the runtime behavior of these packages.
By tracking interactions between their dependencies and the OS file system, \depPrune effectively identifies both direct and indirect bloated dependencies.
% RQ 1: We detect bloated dependencies
% dataset
We evaluate \depPrune on a novel dataset of \nbPackages real-world, well-tested \commonjs packages.
Our analysis reveals that \ratioBloatedDeps of the runtime dependencies are bloated, underscoring the widespread prevalence of dependency bloat in \commonjs packages, particularly contributed by indirect dependencies.
% RQ 2: direct bloated dependencies
Meanwhile, we highlight the importance of bloated direct dependencies and the cascading benefits of removing them, in order to  eliminate a substantial number of associated indirect dependencies.
% RQ 3: Comparative study
We further validate \depPrune through a comparative study with the state-of-the-art dynamic coverage-based analysis and static analysis techniques.
Our findings reveal that \depPrune outperforms these techniques by achieving higher accuracy, avoiding the misclassifications due to the limitations of static analysis, and addressing challenges posed by dynamic and non-linear execution patterns in \javascript.

% enhancing the accuracy of debloating \commonjs applications through coverage-based methods.
% overall findings
Our study highlights an opportunity to further enhance code quality by leveraging trace-based methods like \depPrune to improve the accuracy of dependency debloating in dynamic languages like \commonjs.
In the future, we wish to investigate how dependency debloating can assist developers by reducing notification fatigue, raising awareness about the nebraskan package maintainer \citep{Munroe20} and enhancing overall development efficiency.
%%=============================================================%%
\section{Acknowledgments}
This work was supported by Vetenskapsrådet (Swedish Research Council) and the Knut och Alice Wallenbergs Stiftelse. We gratefully acknowledge the funding received from Vetenskapsrådet, with Benoit Baudry as the grant recipient, and the support from the Knut och Alice Wallenbergs Stiftelse, also with Benoit Baudry as the grant recipient.
%%=============================================================%%

\bibliographystyle{plainnat}
\bibliography{main}
\end{document}